\documentclass[article,nojss]{jss}
\shortcites{Venkatesh_Strauss_Grotegut_2020,Say_Chou_Gemmill_2014,glmmTMB_RJ,chernozhukov2018double,Myers_2011,Powers_Qian_Jung_2018,pkg:glmmTMB}

\let\maxwidth\textwidth

\usepackage{amsfonts,amstext,amsmath,amssymb,amsthm}
\usepackage{accents}
\usepackage{rotating}
\usepackage{nicefrac}
\usepackage{subfigure}
\usepackage{floatrow}
\usepackage{orcidlink}

\usepackage{changes}

\newcommand{\indep}{\perp \!\!\! \perp}

\newcommand\norm[1]{\left\lVert#1\right\rVert}

\usepackage{booktabs}
\usepackage{accents}
\newcommand{\ubar}[1]{\underaccent{\bar}{#1}}

\newcommand{\rZ}{Z}
\newcommand{\rY}{Y}
\newcommand{\rX}{\mX}

\newcommand{\ry}{y}
\newcommand{\rx}{\xvec}

\newcommand{\Xspace}{\mathcal{X}}
\newcommand{\h}{h}

\newcommand{\parm}{\varthetavec}

\newcommand{\ie}{\textit{i.e.}~}
\newcommand{\eg}{\textit{e.g.}~}

\newcommand{\Ex}{\mathbb{E}}
\newcommand{\RR}{\mathbb{R}}

\usepackage{dsfont}
\newcommand{\I}{\mathds{1}}

\DeclareMathOperator*{\argmin}{{arg\,min}}

\DeclareMathOperator{\ND}{N}

\DeclareMathOperator{\PoD}{Po}

\DeclareMathOperator{\BD}{B}
\DeclareMathOperator{\MD}{M}

\def \xvec {\text{\boldmath$x$}}    \def \mX {\text{\boldmath$X$}}

\def \varthetavec     {\text{\boldmath$\vartheta$}}

\author{Susanne Dandl~\orcidlink{0000-0003-4324-4163} \\ LMU Munich \\
	MCML \\
\And Christian Haslinger~\orcidlink{0000-0003-3344-9480} \\
Universit\"atsspital und\\Universit\"at Z\"urich \\
\And Torsten Hothorn~\orcidlink{0000-0001-8301-0471} \\ Universit\"at Z\"urich \\
\AND Heidi Seibold~\orcidlink{0000-0002-8960-9642} \\ IGDORE M\"unchen \\
\And Erik Sverdrup~\orcidlink{0000-0001-6093-1390} \\ Stanford University \\
\And Stefan Wager~\orcidlink{0000-0002-7526-9077} \\ Stanford University \\
\And Achim Zeileis~\orcidlink{0000-0003-0918-3766} \\ Universit\"at Innsbruck}
\Plainauthor{Dandl, Haslinger, Hothorn, Seibold, Sverdrup, Wager, Zeileis}

\title{What Makes Forest-Based Heterogeneous Treatment Effect Estimators Work?}
\Shorttitle{Forest-Based Heterogeneous Treatment Effect Estimators}

\Abstract{
Estimation of heterogeneous treatment effects (HTE) is of prime importance
in many disciplines, from personalized medicine to economics among many others.
Random forests have been shown to be a flexible and powerful approach to
HTE estimation in both randomized trials and observational studies. In
particular ``causal forests'', introduced by \cite{Athey_Tibshirani_Wager_2019},
along with the \proglang{R} implementation in package \pkg{grf} were rapidly
adopted. A related approach, called ``model-based forests'', that is geared
towards randomized trials and simultaneously captures effects of both prognostic
and predictive variables, was introduced by \cite{Seibold_Zeileis_Hothorn_2017}
along with a modular implementation in the \proglang{R} package \pkg{model4you}.

Neither procedure is directly applicable to the estimation of
individualized predictions of excess postpartum blood loss caused by a cesarean section
in comparison to vaginal delivery. Clearly, randomization is hardly
possible in this setup and thus model-based forests lack clinical trial
data to address this question. On the other hand, the skewed and interval-censored postpartum blood loss
observations violate assumptions made by causal forests. Here, 
we present a tailored model-based forest for skewed and
interval-censored data to infer possible predictive prepartum
characteristics and their impact on excess postpartum blood loss caused by a
cesarean section.

As a methodological basis, we propose a unifying view on causal and
model-based forests that goes beyond the \emph{theoretical} motivations
and investigates which \emph{computational} elements make causal forests so
successful and how these can be blended with the strengths of model-based forests.
To do so, we show that both methods can be understood in terms of the same parameters
and model assumptions for an additive model under $L_2$ loss. This theoretical insight
allows us to implement several flavors of ``model-based causal forests'' and
dissect their different elements \emph{in silico}.

The original causal forests and model-based forests are compared with the
new blended versions in a benchmark study exploring both randomized trials
and observational settings. In the randomized setting, both approaches
performed akin.
If confounding was present in the data generating process, we found local centering of the treatment indicator with the
corresponding propensities to be the main driver for good performance. Local centering
of the outcome was less important, and might be replaced or enhanced by simultaneous split selection
with respect to both prognostic and predictive effects. This lays the foundation for
future research combining random forests for HTE estimation with other types of
models.

}


\Address{
  Susanne Dandl \\
  Institut f\"ur Statistik, Ludwig-Maximilians-Universit\"at M\"unchen, 
  	Munich Center for Machine Learning (MCML), Germany \\

  Christian Haslinger \\
  Klinik f\"ur Geburtshilfe, Universit\"atsspital und Universit\"at Z\"urich, Switzerland \\

  Torsten Hothorn\\
  Institut f\"ur Epidemiologie, Biostatistik und Pr\"avention, Universit\"at Z\"urich \\
  Hirschengraben 84, CH-8001 Z\"urich, Switzerland \\
  E-mail: \email{Torsten.Hothorn@R-project.org}\\

  Heidi Seibold \\
  Institute for Globally Distributed Open Research and Education (IGDORE), \\ M\"unchen, Germany \\

  Erik Sverdrup \\
  Stanford Graduate School of Business, Stanford University, U.S.A. \\
  
  Stefan Wager \\
  Stanford Graduate School of Business, Stanford University, U.S.A. \\

  Achim Zeileis\\
  Faculty of Economics and Statistics, Universit\"at Innsbruck, Austria
}

\begin{document}

%% trick LaTeX to put title/abstract on first page
\vspace*{-10cm}
\newpage

%related_work.tex

\section{Introduction}

\subsection{Challenges in treatment effect estimation for cesarean sections}
\label{subsec:intro1}

Cesarean section is the most frequent surgical procedure performed in young
and healthy women, with currently one out of three babies in the USA being
born that way \citep{Antoine_Young_2021}.  Short-term
postpartum benefits and the perceived safety of the procedure explain the
increase in popularity over the last $50$ years, including the rise of
electively performed cesarean sections.  At the same time, maternal
mortality and morbidity increased globally
\citep{WHO_2012,Say_Chou_Gemmill_2014}.  More
recently, adverse long-term effects, including gynecological and obstetrical
complications in mothers as well as potential and controversially discussed 
immune disorders in their children, have
gained attention \citep{Antoine_Young_2021}.  Lack of clinical trial data
directly comparing outcomes of natural births with those following cesarean
sections render characterization and quantification of such effects
challenging.
Postpartum hemorrhage (PPH), defined as blood loss $\ge 500$ mL within
24 hours after delivery by the \cite{WHO_2012}, is a short-term complication 
associated with maternal morbidity and mortality worldwide. The prevalence
of PPH is increasing in industrialized countries \citep[for the USA, see][]{MacDorman_Declercq_Cabral_2016}.

Management of PPH requires
identification of at risk parturients and calls went out to the statistics,
machine learning, and artificial intelligence communities to develop
and evaluate prognostic models \citep{Ende_2022}. Typically, models for
dichotomized PPH prognosis were created aiming
at either women giving birth by vaginal delivery
\citep{Erickson_Carlson_2020,Akazawa_Hashimoto_Katsuhiko_2021} or at women scheduled for a
cesarean section \citep{Kawakita_Mokhtari_Huang_2019}. 
Models trained on data from both modes of delivery are rare, e.g.,
in \cite{Venkatesh_Strauss_Grotegut_2020} the mode of delivery was not
taken into account as risk factor.
Because of the often elective nature of the decision to
undergo cesarean section, a quantification of the \emph{additional} amount of hemorrhaging
caused by surgery is relevant for the decision process, however, such
information is hard to
extract from stratified prognostic models. 
This is true even more considering the possibility of unplanned cesarean
deliveries following attempted vaginal deliveries.
From a statistical perspective,
estimation of a heterogeneous cesarean section effect is non-trivial
for a number of reasons. First, potential risk factors for PPH, such as
age of the mother, estimated birth weight, gestational age, previous PPH, suspected placental
disorders, or multifetal pregnancy
might have an impact on both the decision to undergo a cesarean section (treatment) 
and postpartum blood loss (outcome). Randomization of mode of delivery is impossible
and thus effects have to be estimated from observational data. Second,
it is hard to obtain exact measurements of postpartum blood loss in the
often hectic environment of a delivery ward, and thus imprecise
assessments via interval-censored observations are only available. Third,
one has to expect a high level of skewness and extreme values in blood loss
measurements, rendering strong distributional assumptions questionable.
Last, the association of prognostic factors and blood loss is expected to
be complex, including nonlinear and interaction terms.

\subsection{Heterogeneous treatment effect estimation and random forests}

\label{subsec:intro2}
In the statistical literature, methods for the estimation of such heterogeneous treatment effects (HTEs) from
randomized trials or observational studies has been receiving a lot of attention during the past
decade, triggered by an increasing demand from personalized medicine and the
need for refined methods in causal inference. In particular, different variations
of random forests \citep{Breiman_2001} have been suggested for HTE estimation,
and seem promising candidates for addressing the statistical challenges we
are facing here. Random forest variants for HTE estimation
can be roughly grouped in two classes.

The \emph{first class} of methods employs random forests to estimate the
expected outcomes given covariates separately in the treatment groups.  
The conditional average treatment effect (CATE) 
then corresponds to the difference in estimated mean factual and counterfactual
outcomes. Notably, the virtual twins method \citep{Foster_Taylor_Ruberg_2011}
has adopted this approach using random forests.
%% regresses the observed outcome against covariates and treatment
%% using a random forest.  The factual and counterfactual
%% outcomes are obtained using the original data points and virtual twins with
%% reverted treatment indicators.
Improvements can be obtained by additionally considering
treatment-covariate-interactions or fitting separate (synthetic) forests for each treatment group
\citep{Foster_Taylor_Ruberg_2011, Dasgupta_Szymczak_Moore_Bailey_2014, Ishwaran_Malley_2014}.
Moreover, \cite{Lu_Sadiq_Feaster_Ishwaran_2018} proposed a bivariate imputation
approach which uses a bivariate splitting rule \citep{Ishwaran_Kogalur_2008,
	Tang_Ishwaran_2017} that simultaneously considers the expected outcome under
both treatments.
%% In each step, a forest is grown on data with the observed
%% and unobserved outcomes where the latter is imputed by a previously fitted
%% forest.  The first forest is grown only on observed outcomes.  The CATE 
%% corresponds to the difference between the factual and counterfactual
%% outcomes, one of which is the original observed response and the other was
%% estimated.
In a more general setup,
\cite{Kuenzel_Sekhon_2019} introduced X-learners, a class of meta-algorithms
which build
upon any supervised/regression algorithm including random forests, Bayesian
regression trees \citep[BART,][]{Chipman_2010, Hill_2011, Starling_Murray_Lohr_2021}, or neural
networks.
%% X-learners consist of three steps: the estimation of separate
%% conditional mean functions for each treatment group, the estimation of
%% separated imputed treatment effects using the estimated outcomes of step 1,
%% and the derivation of conditional average treatment effect using a weighted
%% average of the imputed effects of step 2.
Most forest methods were initially developed for randomized controlled trials and 
have later been adapted to be more robust to confounding. 
For example, the pollinated transformed outcome forests
of \cite{Powers_Qian_Jung_2018} build a single forest on propensity score
weighted outcomes instead of the original outcomes to account for confounding. 

The subject of this paper is the \emph{second class} of random forest-type
algorithms aiming at the \emph{direct} estimation of HTEs in a model-driven
way.  Two such approaches, ``causal forests''
\citep{Athey_Tibshirani_Wager_2019} and ``model-based forests''
\citep{Seibold_Zeileis_Hothorn_2017}, have
recently been proposed. 
%Although the original publications might leave the reader with the impression of two
%rather distinct methodologies being proposed, causal forests and model-based
%forests share common foundations, which we shall excavate in the following. 
%Apart from the many similarities, though, a key difference between these two types
%of forests is their respective strategy to deal with nuisance prognostic effects.
%
``Causal forests'' by \citet{Athey_Tibshirani_Wager_2019} implement a divide-and-conquer strategy,
also referred to as ``local centering'' or ``orthogonalization'' for
the direct estimation of HTEs from observational data. They first account for the
dependence of both the marginal mean of the outcome and the treatment propensity
on the available covariates. Subsequently, they exclusively focus on the estimation
of the HTEs. In terms of distributional assumptions, causal forests have been
developed for continuous outcomes and corresponding conditional means and the 
squared error loss plays an important role in the motivation of this
algorithm. \cite{Cui_Kosorok_Wager_Zhu_2020} also applied causal forests to survival data
and \cite{Mayer_Sverdrup_2020} discussed strategies to handle missing values.
We note that earlier causal tree and forest algorithms described in \citet{Imbens_Athey_2016} and
\citet{Wager_Athey_2018} do not involve such a local centering step. In this paper, we use the term
causal forests to describe the algorithm from \citet{Athey_Tibshirani_Wager_2019};
see also \citet{Athey_Wager_2019}.
Causal forests are implemented in the \proglang{R} package \pkg{grf} \citep{pkg:grf}.

%% Random forest-type
%% algorithms are popular in this domain, as demonstrated by the rapid adoption
%% of ``causal forests'' by \cite{Athey_Tibshirani_Wager_2019}.  This paper and
%% the accompanying \proglang{R} add-on package \pkg{grf} \citep{pkg:grf} generated more than
%% 500 citations in the first two years after publication.  Going beyond the
%% \emph{theoretical} motivations by \cite{Athey_Tibshirani_Wager_2019}
%% explaining why their procedure allows the estimation of HTEs from
%% observational data, we ask which \emph{computational} aspects make this
%% method successful in this domain.

``Model-based forests'' by \cite{Seibold_Zeileis_Hothorn_2017} simultaneously estimate prognostic effects and HTEs.
They do so by leveraging model-based recursive partitioning
\citep[``MOB'',][]{Zeileis_Hothorn_Hornik_2008}, a technique for learning model trees
in which all relevant parameters are re-estimated in each subset of a tree.
MOB is not a specific model but rather a general framework for model construction
where the adaptation to different types of models often still necessitates
working out the details of parameter interpretation or model assessment, etc.
\cite{Seibold_Zeileis_Hothorn_2015} have adapted MOB to model-based trees
for HTE, working out the details for Gaussian regression models as well as
censored survival models (parametric Weibull model and semi-parametric Cox
model). Subsequently, \cite{Seibold_Zeileis_Hothorn_2017} have extended this
work to model-based forests for HTEs, again working out the details of Gaussian
regression and censored Weibull survival modeling. Other authors have adapted
the general MOB idea to outcome variables on other scales and/or subject to
censoring and truncation, e.g., as in
survival data \citep{Korepanova_Seibold_Steffen_2019},
ordinal data \citep{Buri_Hothorn_2020},
generalized mixed models \citep{Fokkema_Smits_Zeileis_2017},
or transformation models \citep{Hothorn_Zeileis_2021}. So far,
model-based forests have only been developed for HTE estimation based on
randomized trial data.

\subsection{Model-based causal forests for postpartum blood loss}

Neither of the random forest approaches from Section~\ref{subsec:intro2} is 
directly applicable to the estimation of heterogeneous cesarean section effects, 
described in Section~\ref{subsec:intro1}. Our main contribution is therefore a novel 
random forest model that combines the strengths of the existing methods to tackle the 
challenges in the cesarean section data. We approach this problem by first studying 
the similarities and differences between causal forests and model-based forests 
theoretically and empirically. In a second step, we identify the key drivers for 
good HTE estimation performance in observational data on the one hand and for 
asymmetric and potentially interval-censored outcomes on the other hand. Lastly, 
we derive and apply the novel ``blended'' HTE random forest for PPH by combining 
the elements identified as being instrumental.

Given that both causal forests and model-based forests encompass additive models
under $L_2$ loss, we adopt this modeling framework to investigate the specific elements
that explain both the success of causal forests for observational studies and the
flexibility of model-based forests for randomized trials. Specifically, the question
of how the disparate strategies for handling the prognostic and confounding effects
differ -- or how they can be combined -- is of both theoretical and practical interest.
For obtaining some answers to this question,
we employ the modular computational toolbox for tree induction and forest inference
in the \proglang{R} package \pkg{model4you} \citep{Seibold_Zeileis_Hothorn_2019} 
which allows to ``mix \& match'' the elements of both model-based and causal forests.

The results lay the foundation for future research that further expands potential
synergies in HTE estimation using \textit{model-based causal forests} by blending
model-based and causal forests to leverage the strengths of both approaches.
To demonstrate this in practice, we investigate the effect of cesarean section on
postpartum blood loss in comparison to vaginal deliveries based on a
prospective observational study from Switzerland. 
In this application, there is a need for a model-based approach that
can deal with the skewed outcome distribution which is also interval-censored due
to the lack of precise measurement techniques.  Thus, we 
showcase a model-based transformation forest applicable to this observational
setting. Our contributions here are three-fold: First, we provide a unified
understanding of causal forests and model-based forests for HTE estimation in
Section~\ref{sec:method}. Second, we evaluate why these methods work in different
scenarios and what the key
drivers for good HTE estimation performance in the observational setting are in
Section~\ref{sec:sim}. Last, based on the insights gained theoretically
and empirically, we discuss a novel ``blended''
random forest model in Section~\ref{sec:hypotheses} specifically designed 
for blood loss prediction by pooling key components from causal and model-based forests
(Section~\ref{sec:birth}).

\section{Models and forest algorithms}
\label{sec:method}

In this section, we first outline similarities and differences between causal
forests and model-based forests theoretically, using the basic setup of
regression for real-valued outcomes.  Subsequently, two novel blended
approaches are introduced that adapt HTE estimation with model-based forests
to observational data.

\subsection{The interaction model}

\label{sec:methodlin}

We are interested in the conditional mean of a real-valued outcome
$\rY \in \RR$, given covariates $\rX \in \Xspace$ under a specific binary treatment
or intervention $W \in \{0, 1\}$, corresponding to control vs.~treatment. 
Under the assumptions that a binomial model $W \mid \rX =
\rx \sim \BD(1, \pi(\rx))$ with propensities $\pi(\rx) = \Prob(W = 1 \mid
\rX = \rx) = \Ex(W \mid \rX = \rx)$ describes treatment assignment and residuals are given by
an error term $\sigma \rZ$ with $\Ex(\rZ \mid \rX, W) = 0$ and standard deviation
$\sigma > 0$, the model reads
\begin{eqnarray}
\rY & = & \mu(\rX) + \tau(\rX) W + \sigma \rZ  \label{eq:lm}
\end{eqnarray}
with conditional mean function
\begin{eqnarray*}
	\Ex(\rY \mid \rX = \rx) = \mu(\rx) + \tau(\rx) \pi(\rx) =: m(\rx). %\label{eq:elm}
\end{eqnarray*}
Covariates $\rx$ with impact on the prognostic effect $\mu(\rx)$ are called
\emph{prognostic}, while covariates affecting the treatment effect
$\tau(\rx)$ are called \emph{predictive}.
Treatment assignment is assumed to be non-deterministic, \ie propensity scores
have to be bounded away from zero and one
$$0 < \pi(\rx) = \Prob(W = 1 \mid \rX = \rx) = \Ex(W \mid \rX = \rx) < 1.$$ 
Personalized medicine and causal
inference in general focus on the estimation of the heterogeneous treatment
effect $\tau(\rx)$ and thus on the impact of predictive variables on
treatment success; and accurate estimation of $\tau(\rx)$ is the main
goal of all methods discussed in this paper.

As discussed in \citet{Nie_Wager_2020}, the interaction model \eqref{eq:lm}
is closely connected to a treatment model with potential outcomes
\citep{Imbens_Rubin_2015}, where we posit potential outcomes $Y(0)$ and $Y(1)$
corresponding to the outcome a unit would have experienced without or
with treatment respectively, and assume that we observe $Y = Y(W)$. Then
under unconfoundedness \citep{Rosenbaum_Rubin_1983}
$$(Y(0), Y(1)) \indep W \mid \rX = \rx,$$ 
we can define residuals $\sigma \rZ$ in \eqref{eq:lm} such that the interaction model
is observationally equivalent to the specification using potential outcomes, and
$$ \tau(\rx) = \text{CATE}(\rx) = \Ex(\rY(1) - \rY(0) \mid \rX = \rx) $$ 
can be interpreted as the conditional average treatment effect. We note that
in a uniformly randomized trial, we have $W \indep \{\rX, \, Y(0), \, Y(1)\}$ and
so unconfoundedness is always satisfied, and the propensity scores
$\pi(\rx) \equiv \pi$ are constant by design.

\subsection{Causal forests}

For developing causal forests, \cite{Athey_Tibshirani_Wager_2019} rewrite
Equation (\ref{eq:lm}) as
\begin{eqnarray}
(\rY \mid \rX = \rx)  &=&  m(\rx) - m(\rx) + \mu(\rx) + \tau(\rx) W + \sigma \rZ \nonumber \\
&=&  m(\rx) + \tau(\rx)(W - \pi(\rx)) + \sigma \rZ
\label{eq:lm2}
\end{eqnarray}
which motivates their algorithmic approach of eliminating the marginal mean
$m(\rx) = \Ex(\rY \mid \rX = \rx)$ and propensities $\pi(\rx) = \Ex(W \mid
\rX = \rx)$ first before estimating the heterogeneous treatment effect
$\tau(\rx)$.  This orthogonalization \citep[introduced by][]{Robinson_1988}
is also called ``local centering'' because both outcome $\rY -
\hat{m}(\rx)$ and treatment indicator $W - \hat{\pi}(\rx)$ are centered
before $\tau(\rx)$ is estimated.  This approach leads to more robustness to confounding effects in case of observational data because it regresses out the effect of covariates
$\rX$ on $\rY$ and $W$ \citep{Nie_Wager_2020}.
While in principle any non-parametric regression technique could be applied to
estimate $m(\rx)$ and $\pi(\rx)$, \cite{Athey_Tibshirani_Wager_2019} chose
regression forests. 
%To estimate the individual treatment effect $\tau(\rx)$ of an observation
% $\rx$ in Eq.~(\ref{eq:lm2})
% \citet[][\code{grf::causal\_forest}]{Athey_Tibshirani_Wager_2019} suggest
% to use causal forests that regard heterogeneity in $\tau$ in conjunction
% with local maximum likelihood estimation.

In the second step of causal forests, treatment effects $\tau(\rx)$ in the model 
\begin{equation*}
(\rY \mid \rX = \rx, W = w) = \hat{m}(\rx) + \tau(\rx)(w - \hat{\pi}(\rx)) + \sigma \rZ
\end{equation*}
are then estimated by minimizing the $L_2$ loss 
\begin{eqnarray*}
	\ell_\text{cf}(\tau(\rx)) := \nicefrac{1}{2}\left(\rY - \hat{m}(\rx) - \tau(\rx)(w - \hat{\pi}(\rx))\right)^2
	\label{eq:losscf}
\end{eqnarray*} w.r.t.~$\tau$, the only unknown quantity in this loss function.

Specifically, when splitting a (parent) node, cut-point estimation for causal trees relies first on
estimating a constant treatment effect $\hat{\tau}$ in the parent node minimizing
$\ell_\text{cf}(\tau)$ by solving the score equation
\begin{equation}
s_\text{cf}(\tau) = -\frac{\partial \ell_\text{cf}(\tau)}{\partial \tau} =
(\rY - \hat{m}(\rx) - \tau(w - \hat{\pi}(\rx)))(w - \hat{\pi}(\rx)) = 0
\label{eq:scorecf}
\end{equation}
and second on regressing the resulting score
\begin{eqnarray*}
	s_\text{cf}(\hat{\tau}) = (\rY - \hat{m}(\rx) - \hat{\tau}(w - \hat{\pi}(\rx)))((w - \hat{\pi}(\rx)))
\end{eqnarray*}
on $\rx$ by means of a simple cut-point model. The classical simultaneous
analysis-of-variance (ANOVA) selection of split variable and cut-point is implemented.
Causal forests are robust to confounding because the score equation \eqref{eq:scorecf}
is Neyman-orthogonal in the sense of \citet{chernozhukov2018double}, thus enabling it
to accurately target $\tau(\rx)$ even when estimators for the nuisance components
$\pi(\rx)$ or $\mu(\rx)$ may be somewhat imprecise \citep{Nie_Wager_2020}.
Of course, causal forests can be also applied to randomized data, in which case
treatment should be centered by the true randomization probability $\pi$.

%Theoretically all splits in all variables are considered for selecting the best cut-point.

% SW: I didn't understand the point being made here -- happy to discuss if anyone thinks it's important.
%Because this is computationally quite expensive, the software implementation (\proglang{R} package \pkg{grf}) approximates these splitting criteria using the information in the parent nodes \citep[details are given in Section 2.3,][]{Athey_Tibshirani_Wager_2019}.

\subsection{Model-based forests} \label{sec:mob}

In contrast to the marginal model~(\ref{eq:lm}) motivating local centering
in causal forests, model-based forests \citep{Seibold_Zeileis_Hothorn_2017} for real-valued
outcomes are based on a model which, in addition to $\rx$, also conditions
on treatment assignment $W = w$:
\begin{eqnarray}
(\rY \mid \rX = \rx, W = w) = \mu(\rx) + \tau(\rx)w + \sigma \rZ.
\label{eq:lm3}
\end{eqnarray}  
The main difference between causal forests and model-based forests is that
the latter aims to estimate both $\mu(\rx)$ and $\tau(\rx)$ simultaneously,
whereas the former applies local centering in a two-step approach, that is,
treating the prognostic effect $\mu(\rx)$ as a nuisance parameter.  More
specifically, by using model~(\ref{eq:lm3}) instead of model~(\ref{eq:lm2}), $(\mu(\rx), \tau(\rx))^\top$ is
\textit{simultaneously} estimated by minimizing the $L_2$ loss
\begin{equation}
\ell_\text{mob}(\mu(\rx), \tau(\rx)) = \nicefrac{1}{2}\left(\rY - \mu(\rx) - \tau(\rx)w\right)^2
\label{eq:lossmob}
\end{equation} 
w.r.t.~$\mu$ and $\tau$, the two unknown quantities in this loss function.

Model-based forests separate split-variable and cut-point selection in a way inspired by unbiased recursive partitioning
procedures.  Specifically, in each node, constants
$(\hat{\mu}, \hat{\tau})^\top$ are estimated by minimizing
$$\ell_\text{mob}(\mu, \tau) := \nicefrac{1}{2}\left(\rY - \mu - \tau w\right)^2$$
w.r.t~both $\mu$ and $\tau$.  A split variable is selected by a bivariate
permutation test relying on a quadratic test statistic for the
null hypothesis that $\mu$ and $\tau$ are constant and independent of any
split variable $\rX$.  For splitting, the variable is selected that has the
lowest $p$-value.  Afterwards, a cut-point is found by regressing the
bivariate score
\begin{eqnarray} \label{mobscore}
s_\text{mob}(\hat{\mu}, \hat{\tau}) := (\rY - \hat{\mu} - \hat{\tau} w) (1, w)^\top
\end{eqnarray}
on covariates $\rx$ by a simple bivariate cut-point model.  A cut-point is selected as
the point that results in the largest discrepancy between the score
functions in the two resulting subgroups \citep[details are given in
Appendix 2,][]{Seibold_Zeileis_Hothorn_2017}. The core idea of this
tree-induction method originates from unbiased recursive partitioning
\citep{Hothorn_Hornik_Zeileis_2006} and the introduction of multiple model-based
scores \citep{Zeileis_Hothorn_Hornik_2008} in this framework. 
Section~1 in the Supplementary Material~A provides a more detailed comparison of the cut-point selection of model-based forests with causal forests. 

As a
side-effect, heterogeneous treatment contrasts 
$\tau_{2-1}(\rx)$, $\tau_{3-1}(\rx), \dots$, $\tau_{K - 1}(\rx)$ of $K > 2$
treatment groups $W \mid \rX = \rx \sim \MD(K, \pi(\rx))$ from a multinomial
distribution can be estimated by model-based forests. In each node, the
criterion 
\begin{eqnarray*}
	\frac{1}{2}\left(\rY - \mu(\rx) - \sum_{k = 2}^K \tau_{k - 1}(\rx)w_{k - 1}\right)^2
\end{eqnarray*}
is then minimized w.r.t.~$\mu$ and all treatment contrasts $\tau_{k - 1}$ for $k = 2,
\dots, K$ simultaneously.
This allows the comparison of the effects of different treatments or one treatment with various doses to a placebo \citep[application examples could be found in][]{Schnell_Tang_2017, Feng_Zhou_2012, Zanutto_Lu_Hornik_2005}.

\subsection{Aggregation and honesty}

Once multiple trees have been fitted to sub-samples of the data, causal
forests and model-based forests apply the same local maximum likelihood
aggregation scheme based on nearest neighbor weights for the estimation of
heterogeneous treatment effects $\tau(\rx)$
\citep{Hothorn_Lausen_Benner_2004, Meinshausen_2006, Lin_Jeon_2006,
	Athey_Tibshirani_Wager_2019, Hothorn_Zeileis_2021}.  First, nearest neighbor
weights $\alpha_i(\rx)$ are derived from the $B$~trees in a forest fitted to
observations $(\rY_i, \rx_i, w_i), i = 1, \dots, N$.  These weights measure
the relevance of a training observation $i$ for estimating $\tau(\rx)$.  For
a forest with $B$~trees, $\alpha_i(\rx)$ for an observation $\rx$ is equal
to the frequency with which the $i$-th training sample falls in the same
leaf as $\rx$ over all $B$~trees.  In a second step, $\tau(\rx)$ is
estimated using the reweighted training data by minimizing
$$
\hat{\tau}(\rx) = \argmin_{\tau} \sum_{i = 1}^{n} \alpha^\text{cf}_i(\rx) \ell_{\text{cf},i}(\tau)
%\label{eq:persmod}
$$
in causal forests and
$$
(\hat{\mu}(\rx), \hat{\tau}(\rx))^\top = \argmin_{\mu, \tau} 
\sum_{i = 1}^{n} \alpha^\text{mob}_i(\rx) \ell_{\text{mob},i}(\mu, \tau)
%\label{eq:persmodmob}
$$
in model-based forests, where $\ell_{\text{cf},i}$ and $\ell_{\text{mob},i}$
denote the loss for the $i$-th observation and $\alpha^\text{cf}_i$ and
$\alpha^\text{mob}_i$ are the weights obtained from a causal forest and a
model-based forest, respectively.

\cite{Wager_Athey_2018} additionally recommend a sub-sample
splitting technique called honesty: ``a tree is honest if, for each
training example $i$, it only uses the response $Y_i$ to estimate the
within-leaf treatment effect $\tau$ [...] or to decide where to place the
splits, but not both''. 
They empirically and theoretically proved that honesty is necessary to accomplish valid statistical inference. 
This technique is independent of both tree-induction
and forest aggregation and can be applied in both causal forests and
model-based forests.
In the following, we refer to the \emph{adaptive version} of a tree fitting process, when no sample splitting is conducted,
and we refer to the \emph{honest version}, when honesty is performed. 

\subsection{Model generalizations}
\label{subsec:nonnormal}
When heterogeneous treatment effects shall be estimated for an outcome variable $\rY$ that is
not well described by model~(\ref{eq:lm}), adaptations to
both causal forests and model-based forests are necessary. Causal forests rely on reformulations of the corresponding estimation problems such
that the squared error loss can also be applied in other contexts, for
example in survival analysis \citep{Cui_Kosorok_Wager_Zhu_2020}.
For model-based forests, the loss function $\ell_{\text{mob}}$~(\ref{eq:lossmob}) changes from squared error to the
negative log-likelihood of some appropriate model
\citep[see][]{Seibold_Zeileis_Hothorn_2015,Seibold_Zeileis_Hothorn_2017,Korepanova_Seibold_Steffen_2019,Buri_Hothorn_2020,Fokkema_Smits_Zeileis_2017, Hothorn_Zeileis_2021}.

As a simple example, consider count observations $(\rY \mid \rX =
\rx, W = w) \sim \PoD(\exp(\mu(\rx) + \tau(\rx) w))$ from a conditional Poisson distribution. A
``Poisson forest'' for HTE estimation can be implemented by replacing the
squared error loss~(\ref{eq:lossmob}) with the corresponding Poisson
negative log-likelihood
\begin{equation*}
\ell_\text{mob}(\mu(\rx), \tau(\rx)) = \exp(\mu(\rx) + \tau(\rx) w) - (\mu(\rx) + \tau(\rx) w) \rY.
\end{equation*} 
When it is appropriate to assume $\rZ \sim \ND(0, 1)$ with cumulative distribution function $\Phi$, 
the conditional distribution $(\rY \mid \rX = \rx, W  = w) \sim \ND(\mu(\rx) + \tau(\rx) w, \sigma^2)$
is also normal with cumulative distribution function
\begin{eqnarray*}
	\Prob(\rY \le \ry \mid \rX = \rx, W = w) = \Phi\left(\frac{\ry - \mu(\rx) -
		\tau(\rx) w}{\sigma}\right).
\end{eqnarray*}
For an observed interval $\ubar{\ry} < \rY \le \bar{\ry}$,
model-based forests equipped with the negative log-likelihood
\begin{equation*}
\ell_\text{mob}(\mu(\rx), \tau(\rx), \sigma) = -\log\left(\Phi\left(\frac{\bar{\ry} - \mu(\rx) - \tau(\rx) w}{\sigma}\right) - 
\Phi\left(\frac{\ubar{\ry} - \mu(\rx) - \tau(\rx) w}{\sigma}\right)\right)
\end{equation*} 
allows us to implement a variant of model-based forests applicable to
imprecise interval-censored observations.
In a Tobit model, this is the negative log-likelihood contributed by an observation
$(-\infty, 0]$ left-censored at zero \citep[][equation~(2.1)]{Schlosser_2019}. A
similar likelihood, however without the strict normal assumption, will be
introduced for interval-censored blood loss in Section~\ref{subsec:trafobase}.
In this sense, model-based forests can be understood as a conceptual and
computational framework for method construction, rather than a model with a special domain of application.

\section{Strategies and research questions for blended approaches}
\label{sec:hypotheses}

When applied to data well-described by the additive model~(\ref{eq:lm}) 
in the randomized setting, the principles
underlying causal forests and model-based forests are conceptually the same,
the only difference is that causal forests follow a sequential two-step
approach and model-based forests implement a simultaneous approach
to parameter estimation.  We are now interested in assessing the impact of
implementation details in causal forests and model-based forests on HTE
estimation performance by the two algorithms. The theoretical understanding
from Section~\ref{sec:method} motivates straightforward adaptations to
model-based forests such that the procedure can also be applied to
observational studies. The flexibility of its implementation in
\pkg{model4you} allows to define and evaluate blended estimation
approaches transferring the concept of local centering from causal forests to
model-based forests. Along with these new algorithms, we propose a set of
five research questions which we investigate empirically in Section~\ref{sec:sim}.
An overview of the questions is given in Table~\ref{tab:RQ}.
We begin with the standard implementations of causal forests (cf) and model-based (mob)
forests without centering.  

\paragraph*{RQ 1} How do cf and mob, as implemented in the two
\proglang{R} add-on packages \pkg{grf} (for cf) and \pkg{model4you} (for
mob), compare to each other in randomized and observational settings?

After addressing RQ 1, the question remains if and to what extent local centering inherent in
cf leads to more robustness against confounding effects.  To
answer that we will incorporate orthogonalization in mob as
explained in the following.
Causal forests apply local centering to both the outcome $\rY$ and treatment
indicator $w$, and mob do not center locally at all.  To bring cf and mob
closer, we define a method which applies mob to the model
$$\Ex(\rY \mid \rX = \rx, W = w) = \hat{m}(\rx) + \tilde{\mu}(\rx) + \tau(\rx)(w - \hat{\pi}(\rx)),$$
\ie after centering the treatment indicator $w$ and the outcome $\rY$. 
By using $\tilde{\mu}(\rx)$ instead of $\mu(\rx)$, we emphasize that $\tilde{\mu}(\rx)$ is now the prognostic effect for the \textit{centered} $Y$.

The rationale is to estimate the marginal mean and propensities $\pi(\rx)$ as in
cf first and then apply mob to the centered treatment
$w - \hat{\pi}(\rx)$ and centered outcome $\rY - \hat{m}(\rx)$ to obtain the
prognostic and predictive effect. We call this approach $\text{mob}(\hat{W}, \hat{Y})$. 
The bivariate score function for
mob is changed from~(\ref{mobscore}) to
$$s_{\text{mob}(\hat{W}, \hat{Y})}(\hat{\tilde{\mu}}, \hat{\tau}) := (\rY - \hat{m}(\rx) - \hat{\tilde{\mu}} - \hat{\tau} (w - \hat{\pi}(\rx)) (1, w -
\hat{\pi}(\rx))^\top.$$
In cases where local centering of $Y$ effectively regresses out the effect 
of $\rX$ on $Y$, $\tilde{\mu}(\rx)$ will be close to $0$. 
Since removing $\tilde{\mu}$ leads to the conditional mean function underlying cf
$$ \Ex(\rY \mid \rX = \rx, W = w) = \hat{m}(\rx) + \tau(\rx)(w - \hat{\pi}(\rx)),$$
we call this version ``mobcf''.
Both the outcome and the treatment indicator are centered and only splitting with respect
to scores corresponding to the treatment effect $\tau$ is performed, while
intercept scores are ignored in this process. The only difference between
mobcf and $\text{mob}(\hat{W}, \hat{Y})$ is that simultaneous 
splitting in both the intercept and treatment effect parameters is performed
by the latter, whereas the intercept is ignored in the former.

\begin{table}[t]
	\begin{center}
		\caption{Overview of research questions}
		\begin{tabular}{clll} \hline
			RQ  & Question & Methods & Linear predictors   \\ \hline\noalign{\smallskip}
			1 & Comparison of causal forests   & cf & $\hat{m}(\rx) + \tau(\rx)(w - \hat{\pi}(\rx))$  \\
			& and model-based forests & mob & $\mu(\rx) + \tau(\rx)w $ \\
			\noalign{\smallskip}\hline\noalign{\smallskip}
			2 & Effect of splitting only in $\tau(\rx)$ vs.   & mobcf & $\hat{m}(\rx) + \tau(\rx)(w - \hat{\pi}(\rx))$\\
			& in $\tau(\rx)$ and $\tilde{\mu}(\rx)$ & $\text{mob}(\hat{W}, \hat{Y})$   &   $\hat{m}(\rx) + \tilde{\mu}(\rx) + \tau(\rx)(w - \hat{\pi}(\rx))$     \\ \noalign{\smallskip}\hline\noalign{\smallskip}
			3 & Comparison of causal forests   & cf & $\hat{m}(\rx) + \tau(\rx)(w - \hat{\pi}(\rx))$ \\
			& implemented in grf vs. model4you & mobcf & $\hat{m}(\rx) + \tau(\rx)(w - \hat{\pi}(\rx))$ \\ \noalign{\smallskip}\hline\noalign{\smallskip}
			4 & Effect of locally centering $W$ &  mob$(\hat{W})$  & $\mu(\rx) + \tau(\rx)(w - \hat{\pi}(\rx))$   \\
			& in model-based forests & mob & $\mu(\rx) + \tau(\rx)w $ \\ \noalign{\smallskip}\hline\noalign{\smallskip}
			5 & Effect of additionally centering $Y$ & mobcf & $\hat{m}(\rx) + \tau(\rx)(w - \hat{\pi}(\rx))$  \\
			& in model-based forests centering $W$ & $\text{mob}(\hat{W}, \hat{Y})$ & $\hat{m}(\rx) + \tilde{\mu}(\rx) + \tau(\rx)(w - \hat{\pi}(\rx))$  \\
			& & mob$(\hat{W})$ & $\mu(\rx) + \tau(\rx)(w - \hat{\pi}(\rx))$ \\ 
			\noalign{\smallskip}\hline
		\end{tabular}
		\label{tab:RQ}
	\end{center}
\end{table}

\paragraph*{RQ 2} How does $\text{mob}(\hat{W}, \hat{Y})$ perform compared to mobcf?

The mobcf approach helps us to directly compare the different more technical
aspects, such as variable and split point selection or stopping criteria, of tree
induction implemented in \pkg{grf} and \pkg{model4you}, because it can be seen as 
a re-implementation of cf using the computational infrastructure of the
\pkg{model4you} package.

\paragraph*{RQ 3} How does mobcf perform compared to cf implemented in \pkg{grf}?

Centering the response is straightforward under $L_2$ loss but more
difficult under other forms of the likelihood as discussed in
Section~\ref{subsec:nonnormal}.  The questions arise if and to what extent
solely centering of the treatment indicator $w$ already improves the
estimation accuracy in observational settings.  To answer that we define a
``hybrid approach'' $\text{mob}(\hat{W})$ that applies mob to models parameterized
by $\mu(\rx) + \tau(\rx)(w - \hat{\pi}(\rx))$, \ie after solely centering
the $w$ but not the outcome $\rY$.  The score function for
mob is changed from~(\ref{mobscore}) to
$$s_{\text{mob}(\hat{W})}(\hat{\mu}, \hat{\tau}) := (\rY - \hat{\mu} - \hat{\tau} (w - \hat{\pi}(\rx)) (1, w -
\hat{\pi}(\rx))^\top.$$
\paragraph*{RQ 4} How does solely centering of the treatment indicator ($\text{mob}(\hat{W})$) influence the performance of mob without centering in settings with confounding?

The final research question is whether additional outcome centering
improves upon a forest with treatment centering and simultaneous splits in
prognostic and predictive effects as implemented by mob$(\hat{W})$.

\paragraph*{RQ 5} How does mob$(\hat{W})$ perform compared to mob that center both treatment and outcome (mobcf, and mob$(\hat{W}, \hat{Y})$)?

\section{Empirical evaluation}
\label{sec:sim}

In this section, we provide answers to the research questions defined in Section~\ref{sec:hypotheses}
by evaluating the performance of cf and
mob as well as the different blended versions
in a simulation study for normal outcomes, different
predictive and prognostic effects, and a varying number of observations and
covariates.  The reference implementations in the \pkg{grf} and
\pkg{model4you} \proglang{R} add-on packages were used for the
original cf and mob algorithms. Moreover, the blended approaches
from Section~\ref{sec:hypotheses} are implemented using \pkg{model4you}, \ie by fitting
model-based forests after centering of treatment indicators (mob$(\hat{W})$) and
additionally of outcomes (mob$(\hat{W}, \hat{Y})$ and mobcf, with and without explicitly
accounting for $\mu$, respectively).

\subsection{Data-generating process}
\label{subsec:studynie}

The comparison is based on the study settings of \cite{Nie_Wager_2020}.
The authors proposed four study settings - referred to as Setups~A, B, C and D.
For Setup~A, explanatory variables were sampled by $\rX \sim U([0, 1]^P)$ and for the other three setups they used $\rX \sim N(0, \I_{P\times P})$ -- with $P = \{10, 20\}$
($5$ informative and $P - 5$ noise variables).
Treatment was sampled by $W \mid \rX = \rx \sim \BD(1, \pi(\rx))$ with propensity function $\pi(\rx)$ that
varied among the four considered setups:
\begin{eqnarray*}
	\pi(\rx) = \left\{
	\begin{array}{l}
		\pi_{A}(x_1, x_2) = \max\{0.1, \min\{\sin(\pi x_1 x_2), 1-0.1\}\} \\
		\pi_{B} \equiv 0.5 \\
		\pi_{C}(x_2, x_3) = 1/(1 + \exp(x_2+x_3)) \\
		\pi_{D}(x_1, x_2) = 1/(1 + \exp(-x_1) + \exp(-x_2)).
	\end{array} \right.
\end{eqnarray*}
For Setup~B, probability $\pi \equiv 0.5$ referred to a randomized study.
The conditional average treatment effect function for each setup was
given as
\begin{eqnarray*}
	\tau(\rx) = \left\{
	\begin{array}{l}
		\tau_{A}(x_1, x_2) = (x_1 + x_2)/2 \\
		\tau_{B}(x_1, x_2) = x_1 + \log(1 + \exp(x_2)) \\
		\tau_{C}\equiv 1 \\
		\tau_{D}(x_1, x_2, x_3, x_4, x_5) = \max\{x_1 + x_2 + x_3, 0\} - \max\{x_4 + x_5,
		0\}.
	\end{array} \right.
\end{eqnarray*}
For Setup~C, the treatment effect was constant. The prognostic effects were
defined as
\begin{eqnarray*}
	\mu(\rx) = \left\{
	\begin{array}{l} \mu_A(x_1, x_2, x_3, x_4, x_5) = \sin(\pi x_1 x_2)+ 2(x_3 - 0.5)^2 + x_4 + 0.5 x_5 \\
		\mu_B(x_1, x_2, x_3, x_4, x_5) = \max\{x_1 + x_2, x_3, 0\} + \max\{x_4 + x_5, 0\}\\
		\mu_C(x_1, x_2, x_3) = 2\log (1 + \exp(x_1 + x_2 + x_3))\\
		\mu_D(x_1, x_2, x_3, x_4, x_5) = (\max\{x_1 + x_2 + x_3, 0\} + \max\{x_4 + x_5,
		0\})/2.
	\end{array} \right.
\end{eqnarray*}

Overall, Setup~A has complicated confounding that needs to be overcome
before a relatively simple treatment effect function $\tau(\rx)$ can be
estimated.  In Setup~B, it is possible to accurately estimate $\tau$ without
explicitly controlling for confounding.  Setup~C has strong confounding but
the propensity score function is easier to estimate than the prognostic
effect while the treatment effect is constant.  In Setup~D, the treatment and control arms are unrelated, in the sense that $\mathbb{E}[Y \mid \rX, \, W = 1]$
and $\mathbb{E}[Y \mid \rX, \, W = 0]$ are uncorrelated and there is no benefit to jointly learn
them.

As in \cite{Nie_Wager_2020}, we studied a normal linear regression model
$$ (\rY \mid \rX = \rx, W = w) \sim \ND(\mu(\rx) + \tau(\rx)(w - 0.5), 1),$$
where half of the predictive effect was added to the prognostic effect.

All procedures were applied to $100$ learning samples of size $N \in \{800, 1600\}$
and number of explanatory variables $P \in \{10, 20\}$.  In order to minimize the impact of
different implementation details, cf, mob and the blended versions
were grown with the same hyperparameter options,
see Section~\ref{sec:comp}.  Propensities $\pi(\rx)$ and means $m(\rx)$ were estimated by
\pkg{grf} regression forests for local centering in all forest variants.
For the causal forest, the outcome was always centered by $\hat{m}(\rx)$.
In case of randomized data (Setup~B), the treatment indicator was centered by
$\pi \equiv 0.5$, in all other settings, estimated
propensities $\hat{\pi}(\rx)$ were used.

Performance was assessed by the ability of the methods to estimate the
predictive effect $\tau(\rx)$.  The mean squared error
$\Ex_\mX\{(\hat{\tau}(\mX) - \tau(\mX))^2\}$, evaluated on a test sample of
size $1000$, was used to compare the predictive performance of all candidate
models in the 16 different scenarios.
The results are shown in Figure~\ref{fig:normalBnie}.

The results were also analyzed statistically by means of a normal linear mixed model
with log-link, explaining the estimated mean squared error for
$\hat{\tau}(\rx)$ by a four-way interaction of data generating process,
sample size $N$, dimension $P$, and random forest variant.  We estimated the
mean squared error ratios between cf and mob (RQ 1),
between mobcf and mob$(\hat{W}, \hat{Y})$ (RQ 2), 
between cf and the mobcf approach (RQ 3), 
between mob with centered $W$ (mob$(\hat{W})$) and without (mob) (RQ 4), 
and between mob$(\hat{W})$ and mobcf or mob$(\hat{W}, \hat{Y})$ (RQ 5).
For each simulation run, the model featured a corresponding random
intercept reflecting the paired simulation design.
Simultaneous $95\%$ confidence intervals for the mean squared error ratios are
presented along with the estimates.  For example, the ratio of the mean
squared errors of cf and mob in the first line of Table~\ref{tab:lmeradaptiveb}
was $0.663$ with confidence interval $(0.596, 0.738)$. This is in line with
the performance error of cf being at least $59.6\%$ and at most $73.8\%$
of the performance error of mob, with $66.3\%$ denoting the estimate.
Bold, italic and normal fonts are used to indicate superior, inferior, and
equivalent prediction performance.

\subsection{Results}

%%% <Z>: The green and brown colors are suboptimal for protanope viewers. I
%%% would recommend to use two colors from the Okabe-Ito palette instead. This is
%%% the default palette in palette.colors(). (The colors in the box plots could
%%% also be improved but they are not that crucial to understand the plots.)
%%% </ZH>
\begin{figure}
	
	\includegraphics[width=\maxwidth]{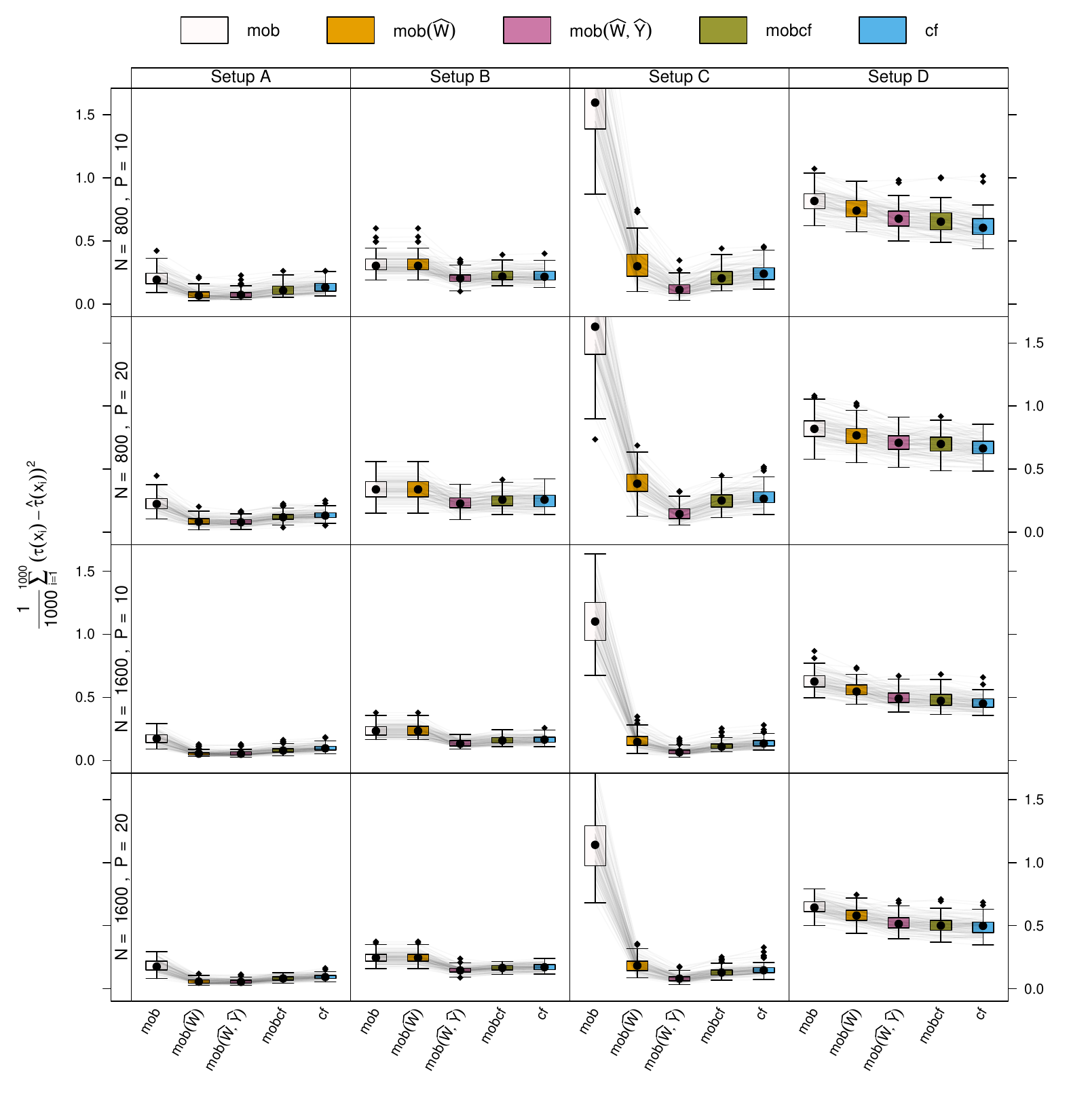} 
	
	\caption{Results for the experimental setups~\ref{subsec:studynie}. Direct comparison of the adaptive versions of causal forests (cf),
		model-based forests without centering (mob), mob imitating causal forests (mobcf), mob with centered $W$ (mob$(\hat{W})$) and additional of $Y$ (mob$(\hat{W}, \hat{Y})$). \label{fig:normalBnie}}
\end{figure}

% TABLES
% Bild model

\begin{sidewaystable}
	\centering
	\caption{Results for the experimental
		setups~\ref{subsec:studynie} for the \textit{adaptive} versions of the
		methods.  Comparison of mean squared errors for $\hat{\tau}(\rx)$ in the
		different scenarios.  Estimates and simultaneous $95\%$ confidence
		intervals were obtained from a normal linear mixed model with log-link.
		Cells printed in bold font correspond to a superior reference (mob in the
		first and fourth columns, mob$(\hat{W}, \hat{Y})$ in the second column, mobcf in the third
		column and mob$(\hat{W})$ in the last column), cells printed in
		italics indicate an inferior reference.  \label{tab:lmeradaptiveb}}
	\scriptsize
	\begin{tabular}{llllrrrrrr}
		\hline
		&& && \multicolumn{6}{c}{Mean squared error ratio}\\
		\cline{5-5}\cline{6-6}\cline{7-7}\cline{8-8}\cline{9-9}\cline{10-10}
		DGP&N&P && \multicolumn{1}{c}{\shortstack{(RQ 1) \\ cf vs. mob \textcolor{white}{$\hat{0}$}}}&\multicolumn{1}{c}{\shortstack{(RQ 2) \\ mobcf vs. mob($\hat{W}, \hat{Y}$)}}&\multicolumn{1}{c}{\shortstack{(RQ 3) \\ cf vs. mobcf\textcolor{white}{$\hat{0}$}}}&\multicolumn{1}{c}{\shortstack{(RQ 4) \\ mob($\hat{W}$) vs. mob}}&\multicolumn{1}{c}{\shortstack{(RQ 5) \\ mob($\hat{W}$) vs. mobcf}}&\multicolumn{1}{c}{\shortstack{(RQ 5) \\ mob($\hat{W}$) vs. mob($\hat{W}, \hat{Y}$)}}\\
		\hline
		
		Setup A&800    &10      && \textit{0.663 (0.596, 0.738)} & \textbf{1.446 (1.206, 1.734)} & \textbf{1.165 (1.017, 1.335)} & \textit{0.392 (0.334, 0.461)} & \textit{0.689 (0.574, 0.826)} &           0.996 (0.806, 1.230)\\
		&       &20      && \textit{0.596 (0.537, 0.662)} & \textbf{1.465 (1.228, 1.747)} &           1.111 (0.972, 1.270) & \textit{0.385 (0.331, 0.446)} & \textit{0.717 (0.604, 0.850)} &           1.050 (0.859, 1.284)\\
		&1600   &10      && \textit{0.575 (0.499, 0.663)} & \textbf{1.458 (1.123, 1.893)} &           1.201 (0.991, 1.455) & \textit{0.324 (0.258, 0.408)} & \textit{0.677 (0.521, 0.881)} &           0.988 (0.727, 1.342)\\
		&       &20      && \textit{0.517 (0.447, 0.598)} & \textbf{1.453 (1.117, 1.889)} &           1.150 (0.944, 1.401) & \textit{0.328 (0.265, 0.407)} & \textit{0.730 (0.567, 0.940)} &           1.061 (0.788, 1.428)\\
		Setup B&800    &10      && \textit{0.707 (0.662, 0.756)} & \textbf{1.099 (1.015, 1.190)} &           0.987 (0.914, 1.065) &           1.000 (0.947, 1.056) & \textbf{1.395 (1.306, 1.491)} & \textbf{1.533 (1.429, 1.646)}\\
		&       &20      && \textit{0.745 (0.701, 0.791)} & \textbf{1.093 (1.018, 1.174)} &           1.001 (0.935, 1.071) &           1.000 (0.951, 1.052) & \textbf{1.345 (1.266, 1.428)} & \textbf{1.470 (1.380, 1.567)}\\
		&1600   &10      && \textit{0.695 (0.635, 0.762)} & \textbf{1.166 (1.036, 1.313)} &           1.034 (0.929, 1.152) &           1.000 (0.929, 1.076) & \textbf{1.487 (1.355, 1.633)} & \textbf{1.734 (1.563, 1.924)}\\
		&       &20      && \textit{0.683 (0.625, 0.746)} &           1.110 (0.992, 1.243) &           1.037 (0.934, 1.152) &           1.000 (0.932, 1.073) & \textbf{1.518 (1.387, 1.662)} & \textbf{1.686 (1.529, 1.859)}\\
		Setup C&800    &10      && \textit{0.148 (0.141, 0.156)} & \textbf{1.693 (1.514, 1.893)} & \textbf{1.150 (1.067, 1.240)} & \textit{0.197 (0.190, 0.205)} & \textbf{1.529 (1.429, 1.636)} & \textbf{2.589 (2.335, 2.870)}\\
		&       &20      && \textit{0.170 (0.162, 0.177)} & \textbf{1.673 (1.520, 1.841)} & \textbf{1.123 (1.051, 1.199)} & \textit{0.236 (0.229, 0.244)} & \textbf{1.563 (1.474, 1.657)} & \textbf{2.615 (2.395, 2.856)}\\
		&1600   &10      && \textit{0.124 (0.113, 0.136)} & \textbf{1.651 (1.348, 2.023)} & \textbf{1.184 (1.032, 1.359)} & \textit{0.143 (0.132, 0.155)} & \textbf{1.368 (1.201, 1.558)} & \textbf{2.258 (1.868, 2.731)}\\
		&       &20      && \textit{0.131 (0.121, 0.142)} & \textbf{1.573 (1.320, 1.875)} & \textbf{1.166 (1.030, 1.320)} & \textit{0.163 (0.153, 0.174)} & \textbf{1.452 (1.295, 1.628)} & \textbf{2.284 (1.943, 2.684)}\\
		Setup D&800    &10      && \textit{0.756 (0.737, 0.775)} & \textit{0.970 (0.945, 0.996)} & \textit{0.934 (0.909, 0.960)} & \textit{0.917 (0.897, 0.938)} & \textbf{1.133 (1.105, 1.162)} & \textbf{1.099 (1.072, 1.127)}\\
		&       &20      && \textit{0.807 (0.788, 0.826)} &           0.983 (0.959, 1.008) & \textit{0.958 (0.933, 0.982)} & \textit{0.926 (0.906, 0.947)} & \textbf{1.100 (1.074, 1.126)} & \textbf{1.081 (1.056, 1.107)}\\
		&1600   &10      && \textit{0.720 (0.696, 0.744)} &           0.970 (0.936, 1.005) & \textit{0.939 (0.904, 0.974)} & \textit{0.885 (0.859, 0.912)} & \textbf{1.155 (1.116, 1.194)} & \textbf{1.120 (1.083, 1.157)}\\
		&       &20      && \textit{0.763 (0.739, 0.787)} &           0.967 (0.935, 1.001) &           0.982 (0.949, 1.018) & \textit{0.894 (0.869, 0.920)} & \textbf{1.151 (1.114, 1.189)} & \textbf{1.113 (1.078, 1.149)}\\
		\hline
	\end{tabular}
	
\end{sidewaystable}

The results for adaptive forests are presented in
Figure~\ref{fig:normalBnie}.  In Section~2 of the Supplementary Material~A,
we report on the effect of honesty
on predictive error as well as the mean squared differences in performance
to cf for the adaptive and honest versions (Figures~S.~1~and~S.~2).
The statistical analysis of the results is given in Table~\ref{tab:lmeradaptiveb} for the adaptive version of forests and in Table~S.~1 of the Supplementary Material~A for the honest version.

\paragraph*{RQ 1.~mob vs. cf}
In all setups, cf outperformed mob.
Especially in Setup~C, mob was unable to overcome the strong confounding
effect and therefore did not provide accurate estimates for the (constant) treatment effect.

\paragraph*{RQ 2.~mob$(\hat{W}, \hat{Y})$ vs. mobcf}
The mob$(\hat{W}, \hat{Y})$ approach performed better than the mobcf approach in almost all scenarios except for Setup~D.
(However, uncorrelated treatment and control arms rarely occur in reality. All methods had a higher MSE than in the other setups.)
These performance differences suggest that splitting by treatment \textit{and} prognostic effect is beneficial.

\paragraph*{RQ 3.~mobcf vs. cf}
Despite the fundamentally different internal splitting and stopping criteria, the original implementation of cf from package \pkg{grf} had
very similar performance to our re-implementation mobcf from package \pkg{model4you} in Setup~A and B.
In Setup~C with strong confounding, the mobcf approach performed slightly better than cf, while in Setup~D cf performed slightly better.

\paragraph*{RQ 4.~mob$(\hat{W})$ vs. mob}
In case of confounding (Setup~A, C), local centering of $W$ (mob$(\hat{W})$)
significantly improved the performance of mob. In Setup~B without confounding,
both approaches performed equally since mob$(\hat{W})$ is equal to mob applied to $w - 0.5$.

\paragraph*{RQ 5.~Methods centering the outcome (mobcf, mobmob$(\hat{W}, \hat{Y})$) vs. mob$(\hat{W})$}
By centering the outcome $Y$ in addition to the treatment $W$, mob$(\hat{W}, \hat{Y})$ and mobcf performed better than mob$(\hat{W})$ except for Setup~A -- centering the outcome did not further improve the results.
The improvements by additionally centering $Y$ were relatively small for mob compared to the improvements due to centering the treatment $W$ (see RQ 4).

Overall, our results reveal treatment effect
centering (mob$(\hat{W})$) as the most relevant ingredient to
random forests for HTE estimation in observational studies.
If possible, additional centering $Y$ in combination with simultaneous
estimation of predictive and prognostic effects (mob$(\hat{W}, \hat{Y})$) is recommended.

\section{Effect of cesarean section on postpartum blood loss}
\label{sec:birth}

\begin{figure}[t!]
	\centering
	\includegraphics[width=0.6\textwidth, trim=0 15 0 50, clip]{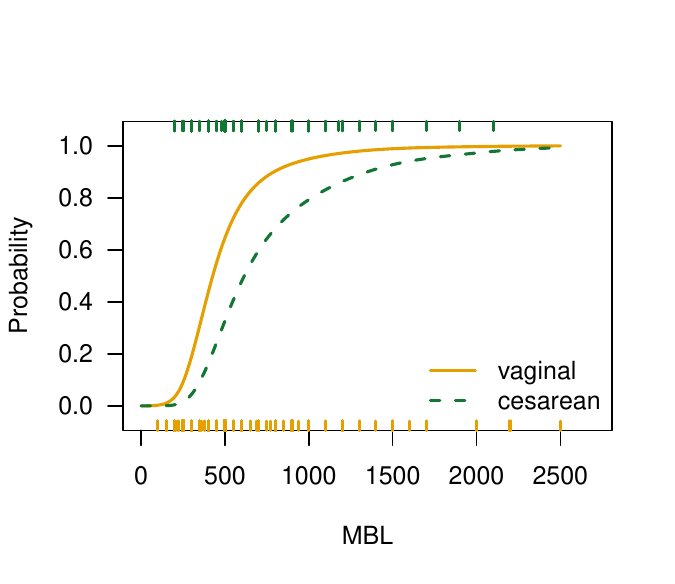}
	\caption{Marginal distribution of measured blood loss (mL) for cesarean section and vaginal delivery. Rugs indicate measured blood loss observations. 
		\label{fig:baseprob}}
\end{figure}

In this section, we discuss
random forest-based HTEs
expressing the additional amount of blood loss explained by prepartum variables, comparing
cesarean sections with vaginal deliveries.  We
analyze data from $1309$ women who participated in a prospective study conducted
from October 2015 to November 2016 at the University Hospital Zurich
\cite[details and data are available from][]{Haslinger_Korte_Hothorn_2020}.  The
outcome is defined as measured blood loss (MBL) in mL and the authors
ensured application of a standardized measurement procedure for all study
participants \citep{Kahr_Brun_Zimmermann_2018}.  For our study, we
removed one outlier observation with a blood loss of $5700$ mL and eight 
observations with missing values for BMI so that a sample of size $N = 1300$ remains.
MBL was recorded as an interval-censored variable, because it is impossible to
exactly determine the amount of blood loss in the sometimes hectic
environment of a delivery ward \citep{Kahr_Brun_Zimmermann_2018}.  Potential
inaccuracies in the measuring process are represented by an interval width
of $50$ mL for blood losses $\le 1$ L and an interval width of $100$ mL when
the mother lost more than one liter of blood.  Measured blood loss can a
priori be considered a positive real and right-skewed variable
(Figure~\ref{fig:baseprob}).  Table~\ref{tab:x} gives a summary of the eight
considered prepartum characteristics ($P = 8$).

\begin{table}[h!]
	\caption{Prepartum characteristics}
	\label{tab:x}
	\begin{tabular}{lll}
		\toprule
		Variable & Description & Range
		\\ 
		\midrule
		GA & Gestational age & 177--297 (days) \\
		AGE & Maternal age & 18--48 (years) \\
		MULTIPAR & Multiparity & no/yes \\
		BMI & Body mass index & 15.4--66 \\
		MULTIFET & Multifetal pregnancy & no/yes \\
		NW & Neonatal weight & 360--4630 (g) \\
		IOL & Induction of labor & no/yes \\
		AIS & Chorioamnionitis & no/yes  \\
		\bottomrule
	\end{tabular}
\end{table}

\newcommand{\mbl}{\text{MBL}}

As the outcome variable MBL is skewed and interval-censored not all
assumptions for causal forests are fulfilled as they estimate a conditional
mean of some continuous outcome optimizing $L_2$ risk.
The extensibility of model-based forests discussed in Section~\ref{subsec:nonnormal}
allows us to take into account the structural assumptions of MBL by substituting $\ell_{\text{mob}}$ in~(\ref{eq:lossmob}) with
the negative log-likelihood of a more appropiate model.
We set up a model-based transformation forest with treatment centering
by combining the mob$(\hat{W})$ approach using local centering
of the treatment indicator within a transformation model.

\subsection{Transformation base model} \label{subsec:trafobase}

The reasoning in Section~\ref{sec:method} is based on the normal linear
model~(\ref{eq:lm3}) and its corresponding likelihood~(\ref{eq:lossmob}) for
absolutely continuous observations. While the latter can easily be adapted to
interval-censored observations, more effort is needed for allowing skewness in
the response distribution.
Adopting a standard normal distribution for the error term $Z$ like in
Section~\ref{subsec:nonnormal}, model~(\ref{eq:lm3}) can be written as a conditional distribution function
\begin{eqnarray*}
	\Prob(\mbl \le y \mid \rX = \rx, W = w) = \Phi\left(\frac{\ry - \mu(\rx) -
		\tau(\rx)w}{\sigma}\right).
\end{eqnarray*}

In this model, symmetry is achieved by a linear transformation of the $\ry$
argument on the probit scale. Replacement of this linear transformation by a
potentially nonlinear one gives rise to transformation models. In
combination with the probit link, this model is a Box-Cox-type linear
regression model that transforms the skewed outcome variable to normality. Instead of
using the traditional Box-Cox power transformation, we estimate a suitable
transformation of MBL by means of a flexible polynomial in Bernstein form \citep{Hothorn_Moest_Buehlmann_2018}.
Ignoring covariates and the local centering of $W$ for a moment, our transformation model describes
the conditional distribution of the positive skewed real variable MBL using
mode of delivery $W$ as treatment indicator for vaginal delivery ($W = 0$)
vs.~cesarean section ($W = 1$):
$$ 
\Prob(\mbl \le y \mid W = w) = \Phi(\h(\ry) - \mu - \tau w).
$$ 
Deviations from normality are captured by the nonlinear transformation
function $\h$ in this model.
Because the transformation function $\h$ contains an intercept term, the
parameter $\mu$ is not identified.  We thus estimate the transformation base model under
the constraint $\mu \equiv 0$.  The intercept function $\h$ varies with the
chosen MBL cut-off $\ry$ and is smooth and monotonically increasing; a
polynomial in Bernstein form of order six was used to parameterize this
function.  The parameter $\tau = \Ex(\h(\rY(1)) - \h(\rY(0)))$ is not 
identical to an average treatment effect on the untransformed scale which could be interpreted 
directly in terms of the original units of the outcome (here blood loss in mL). 
Nevertheless, $\tau$ in our transformation model has an intuitive interpretation 
corresponding to Cohen's d: the units of the treatment effect correspond to standard deviations 
under the normal model.

The parameters of the transformation base model were estimated by minimization of the
negative log-likelihood for an interval-censored observation $(\ubar{\ry}, \bar{\ry}]$
\begin{eqnarray*}
	\ell_\text{Trafo}(\mu, \tau, \parm) & = & -\log(\Prob(\ubar{\ry} < \rY \le \bar{\ry} \mid W = w)) \\ \nonumber
	& = & -\log(\Phi(\h(\bar{\ry} \mid \parm) - \mu - \tau w) - \Phi(\h(\ubar{\ry} \mid \parm) - \mu - \tau w)) %\label{eq:int}
\end{eqnarray*}
where all parameters, including $\parm$ for the transformation
function, are estimated in each node. A parameterisation of $\h$ in terms of
a polynomial in Bernstein form $\h(\cdot \mid \parm)$ ensures uniform convergence to any continuous unknown
transformation function $\h$ on some interval by Weierstrass' approximation
theorem \citep{Farouki_2012}.

\subsection{Personalized transformation model}

The results of Section~\ref{sec:method}--\ref{sec:sim} motivate the
application of model-based forests to a Box-Cox type transformation model
for the estimation of HTEs of cesarean sections on PPH.  The transformation
base model provides skewness and interval-censoring,
whereas the locally centered treatment indicator controls for
potential confounding.  In more detail, we used a mob$(\hat{W})$ forest in combination with
the transformation base model, \ie with local centered treatment
indicator $\hat{w}$, to compute personalized treatment effects $\tau(\rx)$
and prognostic effects $\mu(\rx)$ of the model
\begin{equation}
\Prob(\mbl \le y \mid \rX = \rx, W = w) = \Phi(\h(\ry) - \mu(\rx) - \tau(\rx) (w - \hat{\pi}(\rx))).
\label{eq:persmodblood}
\end{equation}
As in the simulation study, a regression forest was applied to estimate propensities $\pi(\rx)$.
We only used locally centered propensities because the empirical results of
Section~\ref{sec:sim} showed that centering $W$ was the main driver for good performance in observational settings. 
Furthermore, while centering $W$ is straightforward for the transformation model at hand, implementing centering on the outcome $Y$ is less clear.

\begin{figure}[t!]
	\centering
	\includegraphics[width=0.6\textwidth]{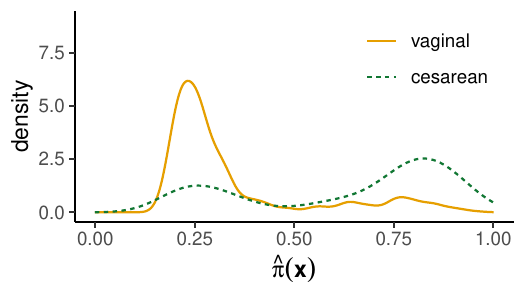}
	\caption{Estimates of propensity scores $\pi(\rx)$ returned by the regression forest for orthogonalization of the treatment indicator} \label{fig:pihat}
\end{figure}

Figure~\ref{fig:pihat} shows that the distribution functions of $\hat{\pi}(\rx)$ for each
treatment group greatly differ. This indicates that prepartum characteristics indeed influence the mode
of delivery and that the treated and control group are dissimilar with respect to these characteristics. 

We first fitted the transformation base model without covariates but
with propensity-centered mode of delivery to estimate a constant effect
adjusted for potential confounding. The corresponding
effect $\hat{\tau}$, \ie the marginal Cohen's
d, was $0.823$ ($\mathrm{CI}_{0.95} = (0.686, 0.959)$), indicating that
women giving birth by cesarean section have a higher postpartum blood loss
compared to women giving birth by vaginal delivery.

The model-based transformation forest was fitted with the same hyperparameter settings as in the 
simulation study~(Section~\ref{sec:comp}). 
We did not adjust the hyperparameters because random forests have been shown to be insensitive to hyperparameter changes \citep{Probst_Boulesteix_2021}.
Figure~S. 3 in the Supplementary Material~A demonstrates this for the mtry parameter -- the number of chosen variables per split. 
We only analysed the mtry parameter since \citet{Probst_Wright_2019} found that the ``mtry parameter is most influential [...]'' while  ``[s]ample size and node size have a minor influence on the performance [...]''.

\begin{figure}[t!]
	\centering
	\includegraphics[width=0.6\textwidth]{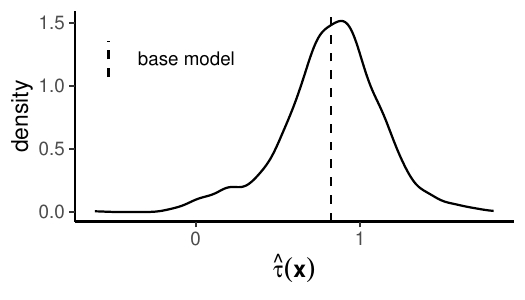}
	\caption{Kernel density estimates of the personalized treatment estimates of the
		model-based transformation forest.
		The dashed line presents the estimated effect of the transformation base model.
		\label{fig:perseff}}
\end{figure}

Figure \ref{fig:perseff} depicts the distribution of the estimated out-of-bag (OOB) heterogeneous treatment effects $\hat{\tau}(\rx)$ of cesarean section compared to vaginal delivery. The distribution is
unimodal and slightly left-skewed.
For almost all births, a cesarean section increases the risk for higher blood losses compared to vaginal delivery. 
For comparison, the average treatment effect of $\hat{\tau} = 0.823$ of the
transformation base model is included.

The interval-censored negative log-likelihood of the transformation base model was $3613.972$. The
model-based transformation forest improved upon this, yielding a likelihood of $3413.989$
(estimated in-bag to make it comparable to the transformation base model).

\begin{figure}[t!]
	\subfigure[Gestational Age]{
		\includegraphics[width=0.23\textwidth]{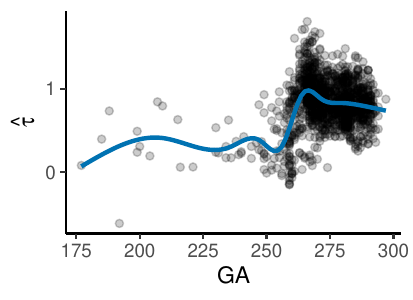}}
	\subfigure[Maternal Age]{
		\includegraphics[width=0.23\textwidth]{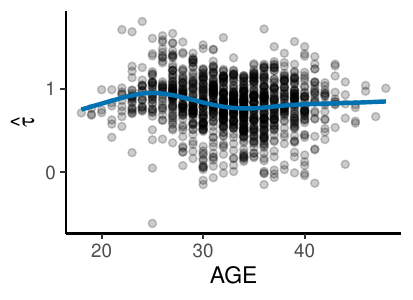}}
	\subfigure[Multiparity]{
		\includegraphics[width=0.23\textwidth]{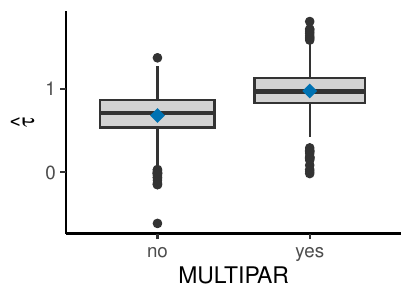}}
	\subfigure[BMI]{
		\includegraphics[width=0.23\textwidth]{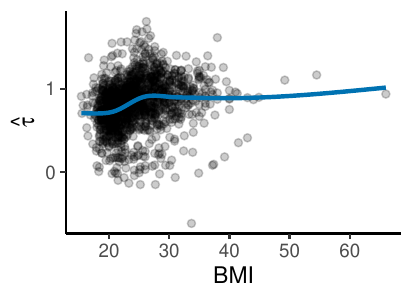}}
	\subfigure[Multifetal]{
		\includegraphics[width=0.23\textwidth]{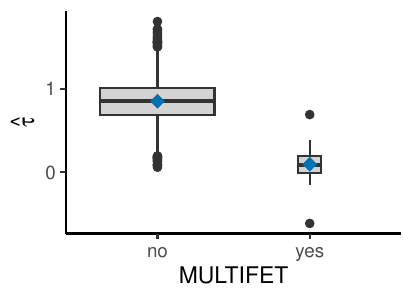}}
	\subfigure[Neonatal Weight]{
		\includegraphics[width=0.23\textwidth]{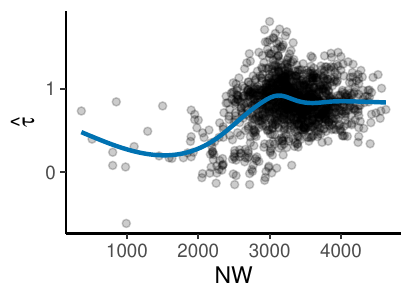}}
	\subfigure[Induction of Labor]{
		\includegraphics[width=0.23\textwidth]{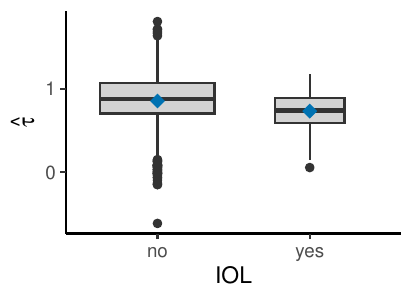}}
	\subfigure[Chorioamnionitis]{
		\includegraphics[width=0.23\textwidth]{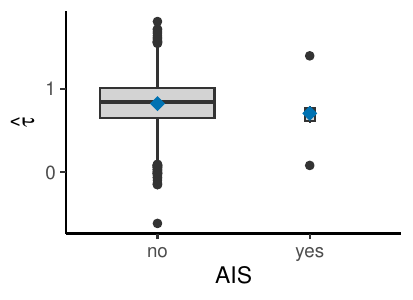}}
	\caption{Dependency plots of the individual treatment effects calculated by the
		model-based transformation forest. Values $\hat{\tau}>0$ mean that cesarean section increases the blood loss compared to vaginal delivery. Lines and diamond points depict (smooth conditional) mean effects.}
	\label{fig:dependplot}
\end{figure}

\begin{figure}[t!]
	\subfigure[Gestational Age]{
		\includegraphics[width=0.2\textwidth]{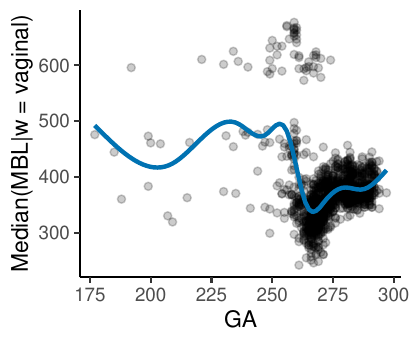}}
	\subfigure[Maternal Age]{
		\includegraphics[width=0.2\textwidth]{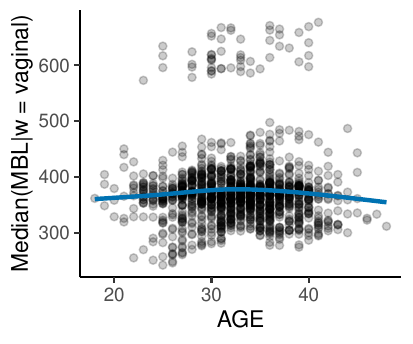}}
	\subfigure[Multiparity]{
		\includegraphics[width=0.2\textwidth]{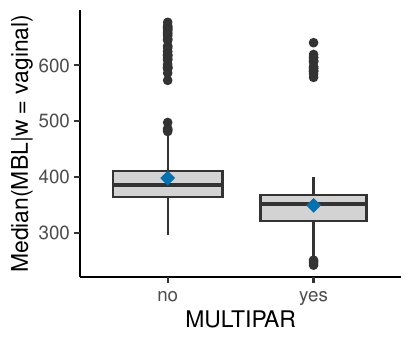}}
	\subfigure[BMI]{
		\includegraphics[width=0.2\textwidth]{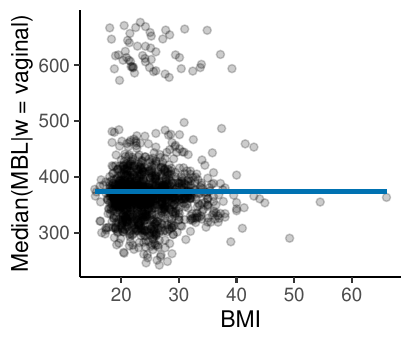}}
	\subfigure[Multifetal]{
		\includegraphics[width=0.2\textwidth]{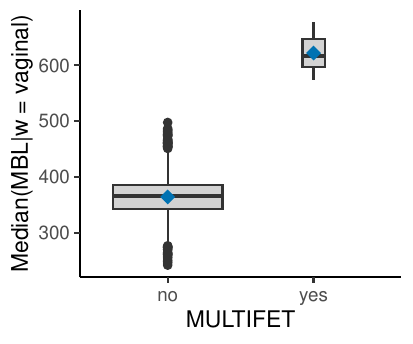}}
	\subfigure[Neonatal Weight]{
		\includegraphics[width=0.2\textwidth]{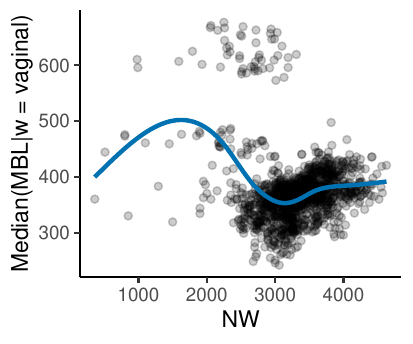}}
	\subfigure[Induction of Labor]{
		\includegraphics[width=0.2\textwidth]{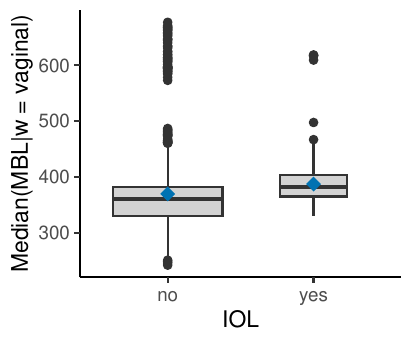}}
	\subfigure[Chorioamnionitis]{
		\includegraphics[width=0.2\textwidth]{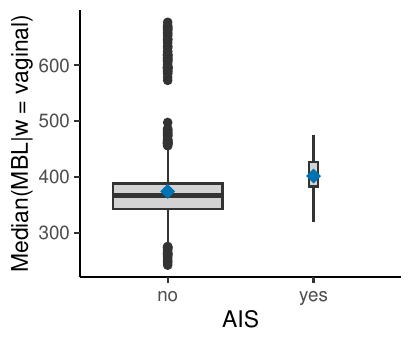}}
	\caption{Dependency plots of median measured blood losses calculated by the
		model-based transformation forest. Higher values mean higher blood loss. Lines and diamond points depict (smooth conditional) means.}
	\label{fig:dependplotalpha}
\end{figure}

\subsection{Dependence plots}

The dependency of the treatment effect $\tau$ on the prepartum variables is
visualized by dependence plots (Figure~\ref{fig:dependplot}).  Scatter
plots are used for continuous covariates and boxplots for categorical
covariates. We also provide mean effects per group for categorical covariates and the smooth conditional mean effect function for continuous covariates. The latter was estimated by a generalized additive model (GAM) with a single smooth term depending on the considered variable.
Births with higher gestational
age, higher neonatal weight and singleton pregnancy have a higher risk
for elevated blood loss due to cesarean section compared to vaginal
delivery. The effect differences were most
pronounced between multifetal and singleton births.  For multifetal
pregnancies, treatment effects are closer to 0 than for singleton pregnancies. 
For a very premature multifetal birth (gestational age of 192
days) of a 25-year-old mother with an elevated BMI of 33.7, a cesarean
section was determined to be most effective ($\hat{\tau} = -0.614$).
Because the distribution of the gestational age (GA) is left-skewed, the curve
of the smoothed conditional mean effects is somewhat erratic.  It might also indicate
that GA was often used as a splitting variable.
While interpreting these results, it should be noted that violations of the unconfoundedness assumption do not seem implausible.

\subsection{Model interpretation and communication}

Interpretation and risk communication in terms of predicted $\hat{\tau}(\rx)$ is difficult
because the effect is defined by Cohen's d on a transformed latent normal scale in
model~(\ref{eq:persmodblood}). However, the model allows conditional quantiles to be computed and
thus information about the conditional MBL distribution for given
prepartum covariates and propensities $\hat{\pi}(\rx)$ can be expressed on the
quantile scale for both modes of delivery.

To assess the prognostic effects on MBL, we computed median measured blood
losses for $W = 0$ (vaginal delivery) given the covariates and propensities.  Figure
\ref{fig:dependplotalpha} indicates that a gestational age of about $270$
days, a birth weight around $3050$ g and singleton births are associated
with small median postpartum blood losses for vaginal deliveries.

The predictive effect of a cesarean section on MBL in such a low-risk group
can be communicated by comparing the MBL distributions under vaginal
delivery and cesarean section.
The median blood loss for a hypothetical
woman in this low-risk group (aged $32.7$ years with a BMI of $24.7$, the mean values
in the study population) is predicted to increase from
$329$ mL (vaginal delivery, $80\%$ prediction interval $209$--$507$ mL) to
$470$ mL (cesarean section, $80\%$ prediction interval $305$--$817$ mL) by our model.
The asymmetric prediction intervals reflect skewness in the MBL
distribution and the wider interval for a cesarean section suggests variance
heterogeneity is captured by the model. The risk of PPH (defined by the
$500$ mL cut-off) is small for vaginal deliveries but substantial under a
cesarean section.

\section{Discussion and outlook}

\subsection{Effects of cesarean sections of postpartum blood loss}

The lives of many of us have been, or will be, impacted by a cesarean
section directly or indirectly.  Empowering women for making an informed
decision, especially in an elective setting, crucially relies on evidence
about the short- and long-term consequences for them and their children
\citep{Antoine_Young_2021}.  Providing an estimate of the individual
predicted excess blood loss caused by a cesarean section, in comparison to a
vaginal delivery, to pregnant women and their obstetricians not only offers
the possibility to decide based on a personalized risk assessment, but has
also the potential to help the overarching goal of reducing the prevalence
of cesarean sections.  The question to perform a cesarean section or not is
less imminent in women with obvious risk factors which make a cesarean
section inevitable (\eg prematurity and multiple fetus pregnancy), but is of
utmost clinical interest in women with a prepartum low-risk profile
(singleton pregnancy at term with normal fetal weight estimation).  To the
best of our knowledge, this is the first study to predict excess postpartum
blood loss in low-risk women.  Our approach of modeling the continuous
blood loss distribution for arbitrary cut-off values is also unique in the
sense that published prognostic models provide risk estimates for events
$\text{MBL} > 500$ mL, or other prespecified cut-off values, only.

Our results were estimated based on data originating from a
prospective study employing a standardized and validated assessment of blood loss under
both modes of delivery. Such efforts can only be
successfully implemented in a controlled setting and hardly apply to
retrospective collections of routine clinical data from multiple study
centers.  However, the detection of smaller but still relevant patterns in
HTEs might require more information than available from the $N = 1300$ study
participants.  The random forest methodology would allow differentiation
between planned and unplanned cesarean sections (Section~\ref{sec:mob}) in a
single model, however, the sample sizes in the present study seem too
limited for such an analysis.  It remains to be seen if refined analyses of
large-scale routine clinical data will provide results similar to those
reported here.

\subsection{Forest-based HTE estimation}

From a statistical perspective,
estimating heterogeneous treatment effects (HTEs) is a difficult task, both
when data from randomized trials and observational studies are analyzed. 
Based on a common theoretical understanding of two strands of random forest
algorithms for HTE estimation, we hypothesized that centering the treatment
with corresponding propensities helps to address confounding.  The empirical
results suggest that this simple modification of the data is instrumental
for the analysis of observational and thus potentially confounded data.

Centering the outcome is equally simple in models for conditional means, but
may be much harder in other models. Empirically, we found that the combination 
of centered treatment and simultaneous split selection (with respect to both 
prognostic and predictive effects) performed at least as well as explicit 
outcome centering. This may seem surprising from a
theoretical point of view, because a nuisance parameter is dealt with in two
completely different ways. Even more interesting is the overall strong
performance of a variant employing both principles at the same time: The
$\text{mob}(\hat{W}, \hat{Y})$ forest is grown on centered outcomes and
treatments and additionally also splits nodes with respect to both prognostic
and predictive effects, leading to a performance at least as well as the
best-performing competitor. Other aspects of tree and forest induction, such as exhaustive search
versus association tests for variable selection, internal stopping criteria
based on sample-size constraints etc., did not explain much variability in
performance.

Based on our current theoretical and empirical understanding of the elements
of both model-based and causal random forests for HTE estimation, we can
make the following recommendations for their application in practice -- especially when
the conditional mean of a numeric outcome captures all relevant aspects:
Data from randomized trials can be analyzed by causal
forests (with outcome centering and known treatment probability $\pi$ for
treatment indicator centering) or model-based forests (with or without
outcome centering) under the intention-to-treat principle.
Under potential confounding, it is important to accurately model
treatment propensities as in causal forests (with outcome and treatment centering).
When combined with treatment centering, model-based forests will lead to
approximately the same results. Additionally centering the outcome
may even offer a small performance gain compared to standard causal forests.

The empirical performances reported in Section~\ref{sec:sim} 
coupled with established asymptotic results for causal random forests with treatment centering 
\citep{Athey_Tibshirani_Wager_2019} and the benign asymptotic behavior of other
ingredients, such as transformation models
\citep{Hothorn_Moest_Buehlmann_2018} or uniform convergence of polynomials in Bernstein form,
suggests favorable asymptotic properties for special flavors of
model-based forests. We leave the presentation of formal results to
future work.

\subsection{Outlook}

The blending of model-based and causal forests
discussed here seems to be
a promising approach for HTE estimation beyond mean regression.
Under potential confounding with binary,
ordinal, count, or survival outcomes, it is easy to combine model-based forests
with treatment centering (mob$(\hat{W})$) following the path outlined in
Section~\ref{subsec:nonnormal}. For example, for a binary
outcome $\rY \in \{0, 1\}$ a logistic regression-based causal forest can
estimate models of the form
$$
\text{logit}(\Prob(\rY = 1 \mid \rX = \rx, W = w)) = \mu(\rx) + \tau(\rx) w.
$$
The HTE $\tau(\rx)$ can then be interpreted as a covariate-dependent log-odds
ratio. In practice, this model can be estimated by package \pkg{model4you}, with
appropriate treatment centering being the only modification necessary
(under the usual assumptions, of course). We leave an in-depth analysis and
evaluation of this principle to future research which should also address the
question of how to achieve outcome centering in such models similar to
mob($\hat{\rY}$, $\hat{W}$).

Finally, going beyond these recommendations and insights, our results are
interesting from two further perspectives. First, the empirical
application to postpartum blood loss in Section~\ref{sec:birth} has shown
that blended model-based causal forests can be tailored to specific setups by 
adapting the underlying loss function.
Second, we empirically demonstrated that two independent implementations of
random forests for HTE estimation performed akin in comparable settings.
This form of external software validation is important in its own right
because the underlying algorithms and implementations are rather complex,
and external validity can only be assessed with the help of an independent
implementation. In case of \pkg{grf} and \pkg{model4you}, past, current, and
future users of these software packages can have higher confidence in
HTEs estimated using either package.

\section{Computational details}
\label{sec:comp}

All computations were performed using \textsf{R} version 4.1.1
\citep{Rcore}, with the following add-on packages: \pkg{grf} \citep{pkg:grf},
\pkg{model4you} \citep{pkg:model4you}, \pkg{trtf} \citep{pkg:trtf}, and
\pkg{partykit} \citep{partykit_MLOSS,pkg:partykit}.

In all empirical experiments, both causal forests and all variants of
model-based forests were grown with \code{M} $= 500$ trees
(\code{model4you::pmforest} default) with minimum node size of \code{node}
$= 14$, number of chosen variables per split \code{mtry} $=P$ and
subsampling (the latter two being \code{causal\_forest} defaults for $P =
10, 20$).  We chose a minimum node size of $14$ because the default of
\code{partykit::ctree_control} (which \pkg{model4you} is based on) is $7$ but we
require this minimum node size for each of the two treatment groups.  For adaptive
forests 50 \% of data were used to build each tree and for honest forests
subsamples were further cut in half (25 \% to determine splits, 25 \% for
estimation, all \pkg{grf} defaults).
To implement local centering of $W$ in case of randomized data for causal forests,
we set \code{W.hat} to 0.5 within \code{grf::causal_forest}.

We used the transformation forest implementation of the \pkg{trtf} package
\citep{pkg:trtf, Hothorn_Zeileis_2021} for fitting the transformation-based forest in
Section~\ref{sec:birth}.

Ratios and confidence intervals presented in Table~\ref{tab:lmeradaptiveb} and
Table~S.~1 (Supplementary Material~A) were computed by generalized
linear mixed models fitted by the \pkg{glmmTMB} package \citep{pkg:glmmTMB}
and post-hoc inference was performed by the \pkg{multcomp} package
\citep{pkg:multcomp}.

We implemented all study settings in a dedicated \proglang{R} package called
\pkg{htesim}. We also included the code and performance results of the empirical study
as well as the code and dataset on postpartum blood loss.
This should facilitate full reproducibility of all findings in this paper.
The package is published on Github: \url{https://github.com/dandls/htesim}.

\section*{Acknowledgments}

Torsten Hothorn received funding from the Swiss National
Science Foundation, with the Grant No.~200021\_184603, Horizon 2020 Research and Innovation
Programme of the European Union under grant agreement number 681094, and is
supported by the Swiss State Secretariat for Education, Research and
Innovation (SERI) under contract number 15.0137.

%\clearpage

\bibliography{m4y,packages,PPH}

\newpage

\renewcommand\thefigure{S.~\arabic{figure}}    
\renewcommand\thetable{S.~\arabic{table}}
\renewcommand\thesection{A.\arabic{section}}
\setcounter{figure}{0}
\setcounter{table}{0}
\setcounter{section}{0}

\section*{Supplementary Material}

\section{Cut-point selection in detail}
\label{ap:cutpoint}
In this section, we compare cut-point selection of model-based forests with causal forests. For ease of exposition, we only consider $p = 1$ covariate.
Our aim is to divide a parent node with $n$ samples into two child nodes.

Model-based forests allow splits both based on the intercept $\mu$ and treatment effect $\tau$ in the model $\rY = \mu + \tau w + \epsilon$, where $\rY$ is the outcome and $w$ is the treatment assignment.
These two can be centered or not without loss of generality, i.e. $\rY_i := \rY_i - \hat \rY_i$ and  $w_i := w_i - \pi(\rX_i)$. Contrary to model-based forests, causal forests only split according to $\tau$. 

We define $W_i$ as the intercept augmented vector $(1 \, \, w_i)$.
We denote the score function for the above model evaluated in the parent node as $\psi$, a $n \cdot 2$ matrix with columns corresponding to $\mu$ and $\tau$.
Let $n_L$ and $n_R$ be the number of samples in the left and right child node, respectively.

\subsection{Model-based forest criterion}
Model-based forests first select a splitting variable using permutation tests before a split point is found. Since we only consider one covariate, we skip this step and continue with the selection of cut points. 
Let $\Sigma \psi_L$ be the sum of the score vector in the left child.
Let $Vh = \frac{1}{n} \sum_{i=1}^{n} \psi^{\otimes 2}$ be a $2 \cdot 2$ weight matrix.
We define $E = n_L \bar \psi$ with $\bar \psi = (\bar \psi_{\mu}, \bar \psi_{\tau})$ as the vector of average scores in the parent node.
With $Z_{\text{mob}} = \Sigma \psi_L - E$ and the weight matrix $V_{\text{mob}} = \left((n n_L / (n - 1) - n_L^2 / (n-1)) Vh \right)^{-1}$ the model-based forest objective is:
$$C_{\text{mob}} = Z_{\text{mob}}' V_{\text{mob}} Z_{\text{mob}}.$$

\subsection{Causal forest criterion}
Causal forests apply CART splitting on pseudo-outcomes $\rho$. The objective is displayed in Equation 5 of \cite{Athey_Tibshirani_Wager_2019}:  
%which for vector valued splits is implemented with the $L_2$ norm and looks like:
$$C_{\text{cf}} = n_L n_R / n^2 \norm{\bar \rho_L - \bar \rho_R}^2,$$
where $\bar \rho_L$ is the average $\rho$ in the left child%(a length 2 vector)
, and likewise for the right child.
The weight value
%$2 \cdot 2$ weight matrix 
is $A_p = \frac{1}{n} \sum_{i=1}^{n} w_i^2$.
%w_i ^{\otimes 2}$, where $\otimes$ denotes an outer product
The $n \cdot 2$ matrix of pseudo-outcomes $\rho$ are then $\rho = \psi_{\tau} A_p^{-1}$.

The criterion $C_{\text{cf}}$ can also be written as a quadratic form similar to model-based forests:
Define $Z_{\text{cf}} = \bar \psi_{\tau,L} - \bar \psi_{\tau,R}$ and $V_{\text{cf}} = n_L n_R / n^2 A_p^{-2}$ with $\bar \psi_{\tau, L}$ and $\bar \psi_{\tau,R}$ as the average scores in the left and right child. Then $C_{\text{cf}} = Z_{\text{cf}}' V_{\text{cf}} Z_{\text{cf}}$ will have the same argmax as above's $C_{\text{cf}}$.

\section{Empirical results for honest forests}
\label{ap:honesty}
\begin{figure}[ht]
	
	\includegraphics[width=\maxwidth]{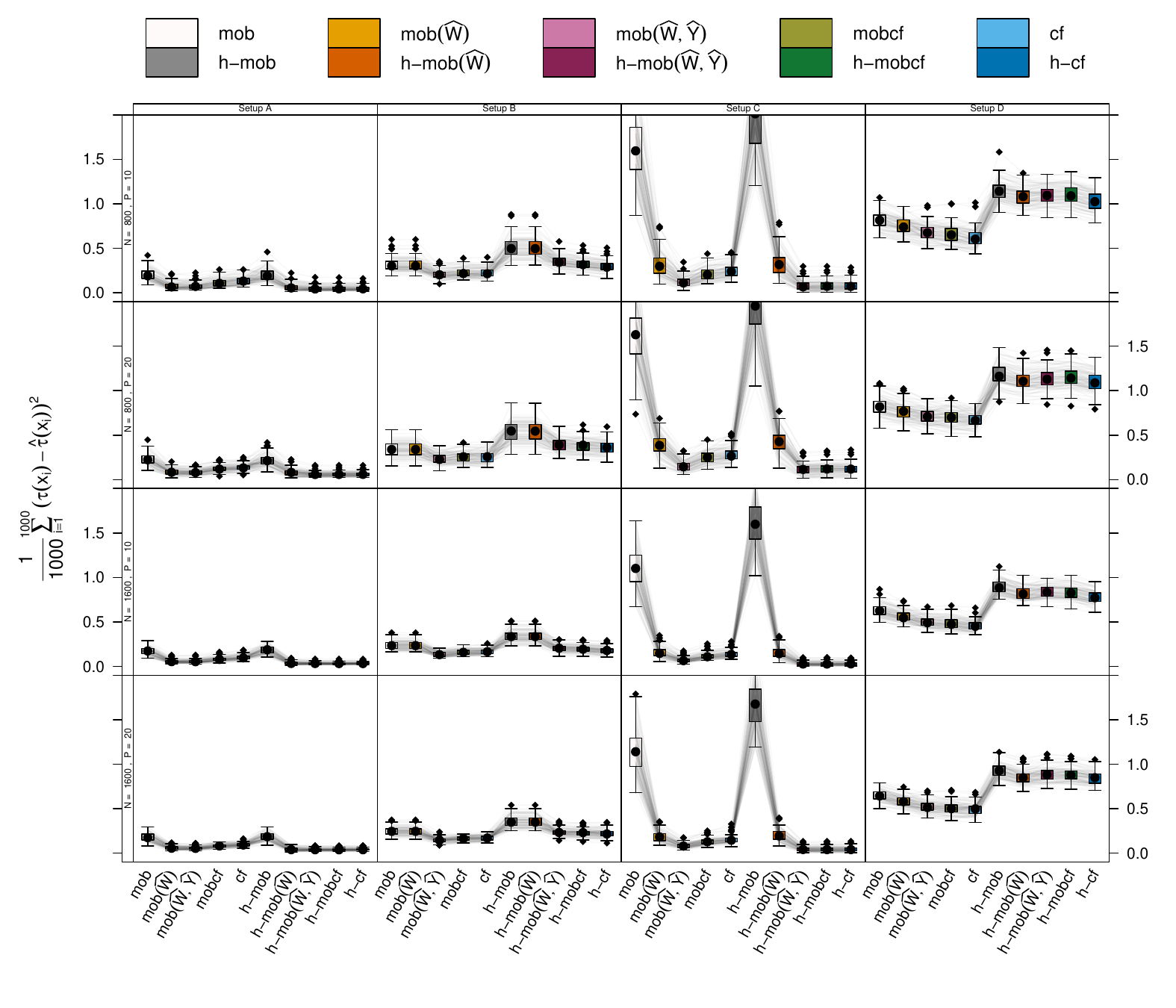} 
	
	\caption{Results for the experimental setups of Section~4. Direct comparison of the adaptive and honest versions of causal forests,
		model-based forests without centering (mob), mob imitating causal forests (mobcf), mob with centered $W$ (mob$(\hat{W})$) and additional of $Y$ (mob$(\hat{W}, \hat{Y})$). 'h-' denotes the honest version of a forest. \label{fig:normalBhonestnie}}
\end{figure}

\begin{figure}
	
	\includegraphics[width=\maxwidth]{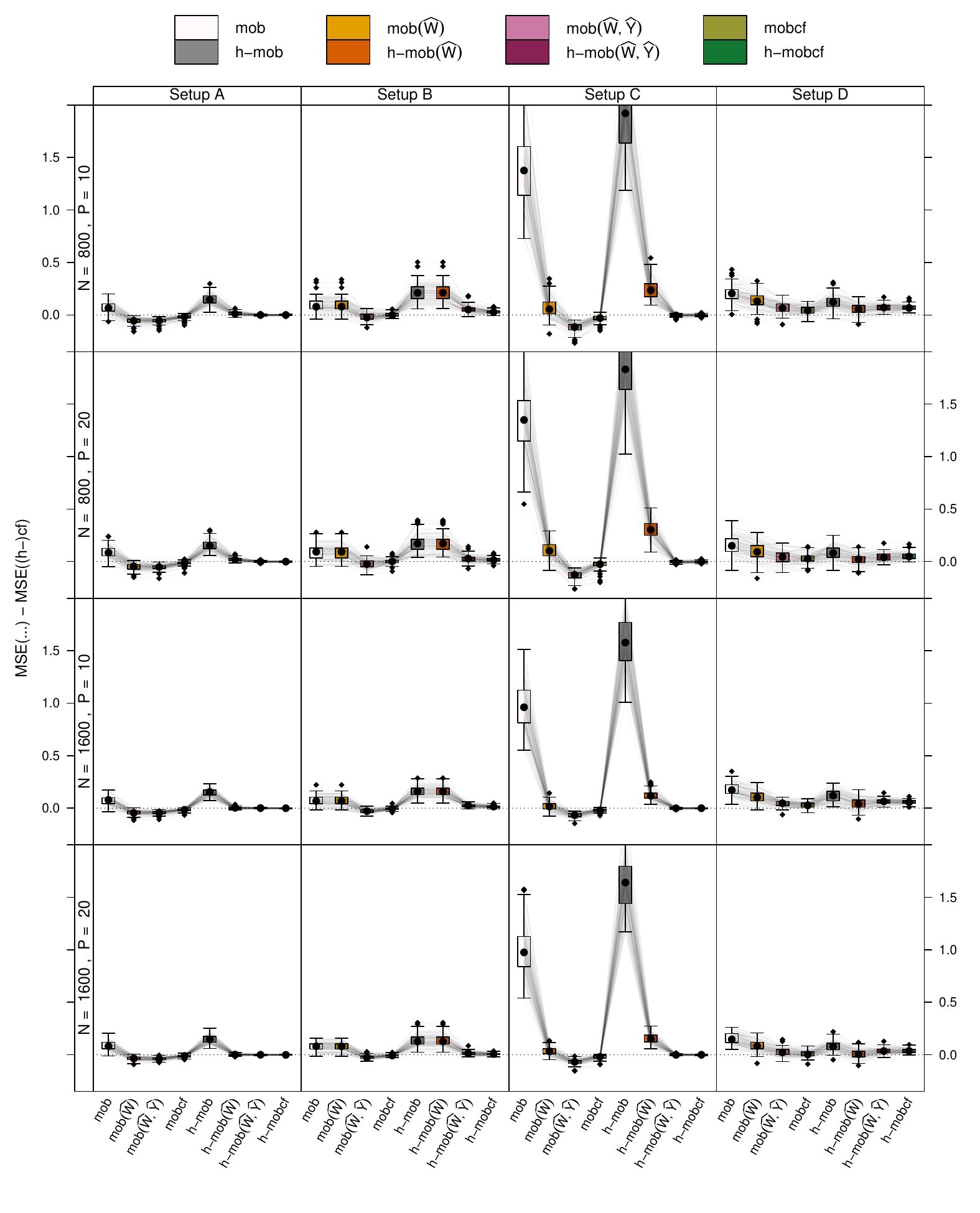} 
	
	\caption{Results for experimental setups of Section~4. Direct comparison of the mean squared differences to causal forests for
		model-based forests without centering (mob), mob imitating causal forests (mobcf), mob with centered $W$ (mob$(\hat{W})$) and additional of $Y$ (mob$(\hat{W}, \hat{Y})$). 'h-' denotes the honest version of a forest. In their adaptive versions, methods were compared to adaptive causal forests, while honest versions to honest causal forests. \label{fig:normalBniesubh}}
\end{figure}

Comparative results of adaptive and honest forests are presented in Figures~\ref{fig:normalBhonestnie} and \ref{fig:normalBniesubh} for the study setting of Section 4.
As for adaptive forests we statistically analyzed honest forests (Table~\ref{tab:lmerhonestnie}).
Rankings of the methods in their honest versions were in line with the results for the adaptive versions.
Most pronounced differences occurred for RQ 2:
While mob$(\hat{Y}, \hat{W})$ performed slightly better than mobcf in their
adaptive versions, they performed akin in their honest versions.
Additional splitting based on the prognostic effect in model-based forests thus had
a smaller impact on performance.
Honesty was beneficial in Setups A and C with strong or complicated confounding. For Setup B, the results differed only slightly in favor of the adaptive versions. For Setup D, honesty worsened the results of all forest approaches.

\begin{sidewaystable}[ht]
	\centering
	\caption{Results for the experimental setups of Section~4 for the \textit{honest} versions of the
		methods.  Comparison of differences in mean squared error for
		$\hat{\tau}(\rx)$ in different scenarios.  Estimates and $95\%$ confidence
		intervals were obtained from a normal linear mixed model with log-link.
		Cells printed in bold font correspond to a superior reference (mob in the
		first and fourth columns, mob$(\hat{W}, \hat{Y})$ in the second column, mobcf in the third
		column and mob$(\hat{W})$ in the last column), cells printed in
		italics indicate an inferior reference.  \label{tab:lmerhonestnie}}
	\tiny
	\begin{tabular}{llllrrrrrr}
		\hline
		&& && \multicolumn{6}{c}{Mean squared error ratio}\\
		\cline{5-5}\cline{6-6}\cline{7-7}\cline{8-8}\cline{9-9}\cline{10-10}
		DGP&N&P && \multicolumn{1}{c}{\shortstack{(RQ 1) \\ cf vs. mob \textcolor{white}{$\hat{0}$}}}&\multicolumn{1}{c}{\shortstack{(RQ 2) \\ mobcf vs. mob($\hat{W}, \hat{Y}$)}}&\multicolumn{1}{c}{\shortstack{(RQ 3) \\ cf vs. mobcf\textcolor{white}{$\hat{0}$}}}&\multicolumn{1}{c}{\shortstack{(RQ 4) \\ mob($\hat{W}$) vs. mob}}&\multicolumn{1}{c}{\shortstack{(RQ 5) \\ mob($\hat{W}$) vs. mobcf}}&\multicolumn{1}{c}{\shortstack{(RQ 5) \\ mob($\hat{W}$) vs. mob($\hat{W}, \hat{Y}$)}}\\
		\hline
		
		Setup A&800    &10      && \textit{0.262 (0.212, 0.324)} &           1.007 (0.752, 1.348) &           1.008 (0.754, 1.347) & \textit{0.336 (0.284, 0.398)} &           1.292 (0.996, 1.677) & \textbf{1.301 (1.001, 1.690)}\\
		&       &20      && \textit{0.284 (0.237, 0.342)} &           1.008 (0.782, 1.299) &           1.012 (0.787, 1.300) & \textit{0.377 (0.327, 0.434)} & \textbf{1.340 (1.073, 1.675)} & \textbf{1.351 (1.080, 1.690)}\\
		&1600   &10      && \textit{0.189 (0.137, 0.261)} &           0.979 (0.622, 1.542) &           1.028 (0.654, 1.616) & \textit{0.202 (0.149, 0.273)} &           1.096 (0.706, 1.700) &           1.073 (0.695, 1.656)\\
		&       &20      && \textit{0.223 (0.171, 0.292)} &           0.991 (0.680, 1.446) &           1.025 (0.705, 1.490) & \textit{0.233 (0.180, 0.301)} &           1.069 (0.740, 1.543) &           1.059 (0.735, 1.527)\\
		Setup B&800    &10      && \textit{0.578 (0.553, 0.604)} & \textit{0.926 (0.883, 0.971)} & \textit{0.915 (0.869, 0.963)} &           1.000 (0.970, 1.031) & \textbf{1.584 (1.520, 1.650)} & \textbf{1.467 (1.411, 1.525)}\\
		&       &20      && \textit{0.677 (0.652, 0.703)} &           0.969 (0.930, 1.010) &           0.962 (0.922, 1.004) &           1.000 (0.971, 1.030) & \textbf{1.422 (1.371, 1.474)} & \textbf{1.378 (1.330, 1.428)}\\
		&1600   &10      && \textit{0.526 (0.491, 0.565)} &           0.957 (0.884, 1.037) &           0.924 (0.849, 1.006) &           1.000 (0.955, 1.048) & \textbf{1.756 (1.644, 1.876)} & \textbf{1.681 (1.577, 1.793)}\\
		&       &20      && \textit{0.621 (0.586, 0.659)} &           0.976 (0.912, 1.046) &           0.973 (0.907, 1.043) &           1.000 (0.957, 1.045) & \textbf{1.565 (1.477, 1.659)} & \textbf{1.528 (1.443, 1.618)}\\
		Setup C&800    &10      && \textit{0.044 (0.039, 0.050)} &           1.046 (0.868, 1.261) &           1.019 (0.850, 1.220) & \textit{0.168 (0.163, 0.174)} & \textbf{3.902 (3.416, 4.458)} & \textbf{4.084 (3.554, 4.692)}\\
		&       &20      && \textit{0.062 (0.057, 0.068)} &           1.053 (0.923, 1.201) &           1.001 (0.881, 1.138) & \textit{0.216 (0.210, 0.222)} & \textbf{3.467 (3.155, 3.810)} & \textbf{3.649 (3.305, 4.029)}\\
		&1600   &10      && \textit{0.020 (0.015, 0.029)} &           1.016 (0.620, 1.664) &           1.024 (0.631, 1.661) & \textit{0.097 (0.090, 0.104)} & \textbf{4.844 (3.402, 6.898)} & \textbf{4.921 (3.437, 7.044)}\\
		&       &20      && \textit{0.028 (0.022, 0.036)} &           0.994 (0.708, 1.396) &           1.010 (0.720, 1.417) & \textit{0.122 (0.115, 0.128)} & \textbf{4.390 (3.429, 5.620)} & \textbf{4.364 (3.414, 5.579)}\\
		Setup D&800    &10      && \textit{0.895 (0.882, 0.909)} &           1.001 (0.987, 1.016) & \textit{0.936 (0.922, 0.950)} & \textit{0.942 (0.929, 0.956)} & \textit{0.985 (0.970, 0.999)} &           0.986 (0.972, 1.000)\\
		&       &20      && \textit{0.929 (0.916, 0.942)} &           1.010 (0.996, 1.024) & \textit{0.955 (0.941, 0.968)} & \textit{0.945 (0.932, 0.958)} & \textit{0.971 (0.957, 0.985)} & \textit{0.980 (0.967, 0.994)}\\
		&1600   &10      && \textit{0.868 (0.851, 0.885)} &           0.992 (0.973, 1.011) & \textit{0.931 (0.913, 0.949)} & \textit{0.915 (0.898, 0.932)} &           0.981 (0.963, 1.000) & \textit{0.973 (0.955, 0.992)}\\
		&       &20      && \textit{0.913 (0.896, 0.929)} &           1.000 (0.982, 1.018) & \textit{0.958 (0.940, 0.975)} & \textit{0.919 (0.903, 0.936)} & \textit{0.964 (0.947, 0.982)} & \textit{0.964 (0.946, 0.982)}\\
		\hline
	\end{tabular}
	
\end{sidewaystable}

\clearpage

\clearpage

\section{Sensitivity of mtry parameter}

\begin{figure}[h!]
	\centering
	\includegraphics[width=.6\textwidth]{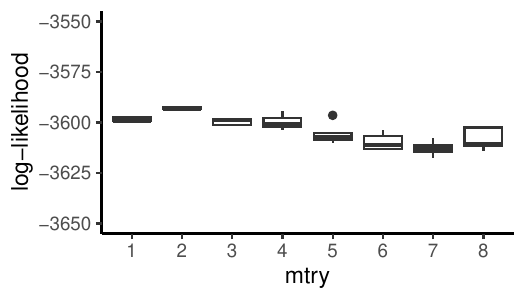}
	\caption{Effect of the mtry parameter on (out-of-bag) log-likelihood of the transformation forest (Section 5). 	Forest fitting was repeated 5 times for each mtry parameter.  All other hyperparameters of the transformation forest were kept at their respective values according to Section~7.}
	\label{fig:mtrymseloglik}
\end{figure}

Sensitivity of the random forest for PPH presented in Section~5 of the main
manuscript was studied with respect to different choices of the main tuning
parameter, \code{mtry} (the number of randomly selected covariates for
split evaluation in each node of the underlying trees). In
Figure~\ref{fig:mtrymseloglik}, the out-of-bag log-likelihoods for
several choices of \code{mtry} are presented, showing an insignificant
amount of variability and thus results can be expected to be quite 
stable with respect to the choice of \code{mtry}.

\end{document}